\documentstyle[epsfig]{mn}
\begin{document}


\title[Observations of HDF South with {\em ISO} - II.]
{Observations of the {\em Hubble Deep Field South} 
with the {\em Infrared Space Observatory} - II. 
Associations and star formation rates} 
\author[R.G. Mann  {\it et al.}]
{ 
Robert G. Mann$^{1,2}$,
Seb Oliver$^{1,3}$, 
Ruth Carballo$^{4,5}$, 
Alberto Franceschini$^6$, 
\newauthor
Michael Rowan--Robinson$^1$, 
Alan F. Heavens$^2$,
Maria Kontizas$^7$, David Elbaz$^8$, 
\newauthor
Anastasios Dapergolas$^9$, 
Evanghelos Kontizas$^9$, 
Gian Luigi Granato$^6$,
\newauthor
Laura Silva$^{10}$,
Dimitra Rigopoulou$^{11}$,
J. Ignacio Gonzalez--Serrano$^{5}$,
\newauthor
Aprajita Verma$^{1,11}$, 
Steve Serjeant$^{1,12}$, 
Andreas Efstathiou$^1$,
Paul P. van der Werf$^{13}$
\\
$^1$Astrophysics Group, Imperial College London, Blackett Laboratory,
Prince Consort Road, London SW7 2BZ\\ 
$^2$Institute for Astronomy, University of Edinburgh, Royal Observatory, Blackford Hill, Edinburgh, EH9 3NJ\\
$^3$Astronomy Centre, School of Chemistry, Physics and Environmental
Science, University of Sussex, Falmer, Brighton, BN1 9QJ\\
$^4$Departamento de Matematica Aplicada y CC, Universidad de Cantabria, Avda. Los
Castros s/n, 39005 Santander, Spain\\
$^5$Instituto de F\'{\i}sica de Cantabria (CSIC-UC), Avda. Los Castros s/n,
39005 Santander, Spain.\\
$^6$Dipartimento di Astronomia --- Universita' di Padova, Vicolo 
dell'Osservatorio 5, I-35122, Padova, Italy\\
$^7$Department of Physics, 
University of Athens, Panepistimiopolis, GR-15783, Zografos, Greece\\
$^8$DSM/DAPNIA/SAp, CE - Saclay, Orme des Merisiers - Bat 709, 91191 
Gif-sur-Yvette Cedex, France\\
$^9$Astronomical Institute, National Observatory of Athens, Lofos Nymfon, 
Thission, P.O. Box 20048, 11810 Athens, Greece\\    
$^{10}$Astrophysics Sector, SISSA, Via Beirut 2-4, 34013 Trieste, Italy\\ 
$^{11}$Max-Planck-Institut f\"ur extraterrestrische Physik, Postfach
1603,85740 Garching, Germany\\
$^{12}$Unit for Space Sciences and Astrophysics, School of Physical Sciences,
University of Kent, Canterbury, Kent, CT2 7NZ\\
$^{13}$Leiden Observatory, P.O. Box 9513,  NL-2300 RA Leiden, The
Netherlands
}
 
\pagerange{\pageref{firstpage}--\pageref{lastpage}}
\pubyear{2002}
\volume{}

\label{firstpage}

\maketitle
\begin{abstract}
We present results from a deep mid--infrared survey of the Hubble Deep Field South (HDF--S) region 
performed at 7 and 15$\mu$m with the CAM instrument on board the {\em Infrared Space Observatory} 
({\em ISO}). We found reliable
optical/near--IR associations for 32 of the 35 sources detected in this field by
Oliver et al. (2002, Paper I): eight of them were identified as stars, one is definitely  an AGN, a 
second seems likely
to be an AGN, too, while the remaining 22 appear to be normal spiral or starburst galaxies.
Using model spectral energy distributions (SEDs) of similar galaxies, we compare methods for estimating
the star formation rates (SFRs) in these objects, finding that an estimator based on integrated 
(3--1000$\mu$m) infrared 
luminosity reproduces the model SFRs best. Applying this estimator to model fits to the
SEDs of our 22 spiral and starburst galaxies, we find that they are
forming stars at rates of  $\sim1 - 100$ $M_{\odot} \; {\rm yr}^{-1}$, with a
median value of $\sim 40$ $M_{\odot} \; {\rm yr}^{-1}$, assuming an Einstein -- de Sitter universe
with a Hubble constant of 50~km~$s^{-1}$~Mpc$^{-1}$, and star formation taking place according to
a Salpeter (1955) IMF across the mass range $0.1-100M_{\odot}$.
We split the redshift range $0.0 \leq z \leq 0.6$ into two equal--volume
bins to compute raw estimates of the star formation rate density,  $\dot{\rho}_*$,  contributed by these sources, 
assuming the same cosmology and IMF as above and computing errors
based on estimated uncertainties in the SFRs of individual galaxies. We compare these results 
with other estimates of $\dot{\rho}_*$ made with the same assumptions, showing them to be consistent with the
results of Flores et al. (1999) from their {\em ISO\/} survey of the CFRS 1415+52 field. However, the 
relatively small volume of our survey means that our $\dot{\rho}_*$ estimates suffer from a large sampling
variance, implying that our results, by themselves, do not place tight
constraints on the global mean star formation rate density.
\end{abstract}

\section{Introduction}\label{intro}

In an accompanying paper (Oliver et al. 2002, Paper I) we described the survey
we made of the Hubble Deep Field South (HDF--S\footnote{see 
{\tt www.stsci.edu/ftp/science/hdfsouth/hdfs.html}}) region using the
LW2 (centred at 6.7$\mu$m) and LW3 (15$\mu$m) filters of the CAM (Cesarsky et al. 1996)
instrument on board the {\em Infrared Space Observatory (ISO)\/} (Kessler et al. 1996).
A prime motivation for this project came from the results of previous {\em ISO\/} surveys,
such as our own survey (Serjeant et al. 1997, Goldschmidt et al. 1997, Oliver et al. 1997, Mann et al. 
1997, Rowan--Robinson et al. 1997) of the
northern Hubble Deep Field (HDF--N, Williams et al. 1996) and that of Flores et al. (1999) in the 
Canada--France Redshift Survey 1415+52 field (Lilly et al. 1995), which revealed an infrared luminosity
density at $0.5 \la z \la 1$ implying a higher star formation rate density at those
redshifts than indicated by previous optical studies (e.g. Lilly et al. 1996, 
Madau et al. 1996, Connolly et al. 1997).  

These studies were facilitated by the availability of
multi--wavelength datasets in those well--studied fields, which made possible the association 
of {\em ISO\/} sources with galaxies for which redshifts had been determined through optical
spectroscopy or for which they could be estimated on the basis of multi--band optical/near--infrared
photometry. 
The same factors favour the study of the HDF--S region, and in this paper we discuss the
association of our {\em ISO\/} sources in the HDF--S with objects in optical and near--infrared 
catalogues of that field, and the estimation of their star formation rates. This process is 
significantly easier than was the case for our survey of the northern HDF--N, since, as emphasised
in Paper I, our {\em ISO\/} dataset in the HDF--S is greatly superior, as a result of an
observational strategy and data reduction procedure both revised significantly in the light of 
developing knowledge of the characteristics of the CAM instrument and of the 
source population it probed. 

The plan of the rest of this paper is as follows. In Section~\ref{sect:associations}, 
we describe the 
existing optical and near--infrared catalogues in the HDF--S region with which we shall
seek associations for our {\em ISO\/} sources, review the likelihood ratio method
used to obtain them and present the results of its application. Section~\ref{sect:sfr_methods}
 compares a
variety of commonly used star formation rate (SFR) estimators, through applying them to
model SEDs, while, in  Section~\ref{sect:sed_sfr}, we apply the best of them to the multi--wavelength
photometric datasets compiled for the galaxies with which we have associated our {\em ISO\/}
sources, to yield estimates of their SFRs, and Section~\ref{sect:madau} uses these results to assess
the contribution of {\em ISO\/}--selected sources to the star formation history of the HDF--S.
Finally, in Section~\ref{sect:discuss}, we discuss the results
of this paper and the conclusions we draw from them. Appendices present mathematical details of
the processes of correcting an SFR estimate to a canonical IMF, and of computing the 
sampling variance in estimates of the SFR density using a lognormal model for the cosmological
density field.

\begin{figure}
\epsfig{file=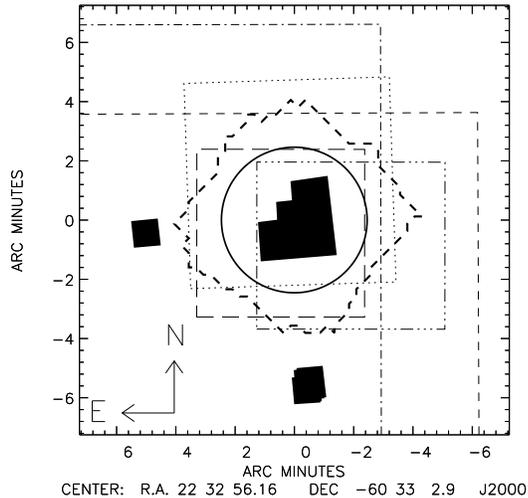,angle=0,width=10cm}
\caption{This figure shows the location of our ISO rasters with
respect to those of other datasets taken in the area. The shaded
regions mark the {\em HST} fields, with the STIS and NICMOS fields
to the east and south of the WFPC2 field, respectively. The 
thick--dashed irregular shape and the solid
circle show, respectively, the maximum extent of our ISO
coverage (the coverage in the two bands differs slightly)
 and the region from which the source catalogues of Paper I were selected. 
The remaining lines show boundaries of
four optical/near-IR surveys discussed in Section \protect\ref{sect:associations},
as follows: (i) dotted line -- AAT prime focus imaging survey of Verma
et al.  (2002); (ii) dashed line -- CTIO BTC survey of Gardner et
al. (1999); (iii) dot-dashed line -- CTIO BTC survey of \protect\cite{Walker 1999};
(iv) dot-dot-dot-dashed line -- ESO EIS optical imaging survey of Da
Costa et al. (1998); and (v) long dashed line -- ESO EIS near-infrared
survey of da Costa et al. (1998).}
\label{fig:fields}
\end{figure}

\section{Optical Catalogues and Associations}\label{sect:associations}

As shown in Figure \ref{fig:fields}, the region within which the final
source catalogue of Paper I was selected is covered by
a number of optical and near--infrared imaging surveys and we have sought
to associate our {\em ISO} {\em sources} with {\em objects} detected in
all of them, using the likelihood ratio method detailed by Sutherland
\& Saunders (1992).

\subsection{Method}

The implementation of the likelihood ratio association technique used
here is essentially identical to that we used for the {\em ISO}--HDF--N
sources in Mann et al. (1997), so we only briefly review it
here. The likelihood ratio, $LR$, for the association of a particular source
with a given catalogue object is defined to be the ratio, 
$p_{\rm true}/p_{chance}$, of the infinitesimal probability of finding 
the true counterpart to the source at the position of the object and
with its flux to the infinitesimal probability of an object with that
flux being found there by chance. Sutherland \& Saunders (1992) show
that it takes the form
\begin{equation}
LR(f,x,y) = \frac {q(f) \cdot e(x,y)}{n(f)},
\label{eq:lr_def}
\end{equation}
where $e(x,y)$ is the probability distribution for positional offsets,
$(x,y)$, between source and object [normalized so that $\int e(x,y)
\; {\rm d}x \; {\rm d}y =1$ with the integral being taken over all
space], $n(f)$ is the surface density of objects per unit interval in
flux, $f$, and $q(f)$ is the probability distribution function for 
an ensemble of sources, measured in the same passband in which the
object catalogue is defined. 

\begin{figure}
\epsfig{file=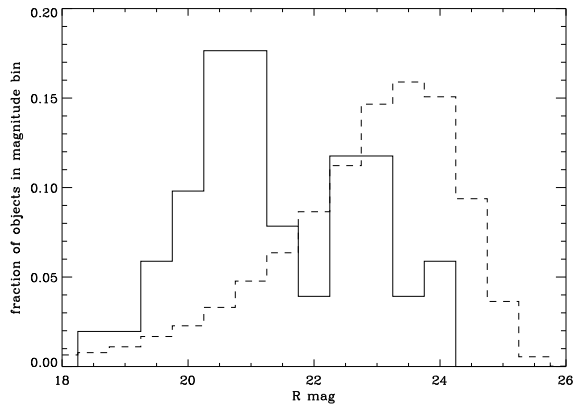,width=8cm}
\caption{A comparison of (solid line) the R band magnitude distribution of the objects within circles of 
radius 6$\arcsec$ around the 30 {\em ISO} sources from Table 8 of Paper I contained within the area of the
Gardner et al. catalogue, and (dashed line) that for the catalogue as a whole, showing the strong excess
of bright objects near {\em ISO} source positions: a two--sided Kolmogorov--Smirnov test gives a probability of
$P \sim 10^{-10}$ that these two are drawn from the same parent distribution.} 
\label{fig:maghistfig}
\end{figure}

The quantity $q(f)$ essentially measures the flux
distribution of the true counterparts of the sources in the object
catalogue, and is unknown unless associations have previously been
found between an ensemble of similar sources and a catalogue of
similar objects. Sutherland \& Saunders (1992) suggested that it may be
estimated in binned form by  subtracting from the flux histogram
of the objects lying near the source positions that for the full
object catalogue. If the
latter is scaled to cover the same area as used to compute the former,
then the resulting quantity is proportional to $q(f)$, exhibiting an
excess of galaxies in the bins corresponding to the fluxes of the true
counterparts of the sources in the object catalogues. Mann et
al. (1997) showed that, when applied to relatively small source
samples, such as in our {\em ISO} surveys of the two Hubble Deep Fields,
this empirical $q(f)$ is very noisy, and they preferred to take $q(f)$
to be a constant, independent of flux. Since their Figure 1 strongly
implied that the {\em ISO--HDF--N} sources were associated with objects
brighter than $I_{814} \simeq 23$ in the catalogue of Williams et
al. (1996), this assumption would have led them to underestimate the
value of $p_{\rm true}$ for bright objects, but, in practice, this was
seen to have no effect on the associations made, so we follow the same
procedure here. The validity of making that same simplifying assumption here 
is illustrated by Fig,~\ref{fig:maghistfig}, where we compare the R band magnitude
distribution near (defined to be within 6$\arcsec$ of) the 30 {\em ISO} sources from 
Table 8 of Paper I contained within the area of the Gardner et al. catalogue, and 
that for the catalogue as a whole. This  shows the strong excess
of bright (R=19--21) objects near {\em ISO} source positions --  a two--sided Kolmogorov--Smirnov test 
gives a probability of $P \sim 10^{-10}$ that these two are drawn from the same parent 
distribution -- which gives us confidence both in the reliability of our source
detections and in our adopting a constant $q(f)$ here.

For each association made between a given {\em ISO--HDF--N} source and a
particular object in the $I_{814}$--band catalogue of Williams et al. (1996), 
Mann et al. (1997) computed the probability, $P_{\rm ran}(I_{814})$,
that a fictitious source, placed at random in the HDF--N,
would have a likeliest association with an object in the $I_{814}$--band
catalogue producing  a likelihood ratio at least as high as that for
the source and object in question. Clearly, $P_{\rm ran}$ combines
information as to the reliability of the source detection as well as
the probability that the given object is the correct counterpart of
the source, but, as discussed below, it provides a useful measure of
the quality of the associations made in this case: a more direct
measure would be available (Sutherland \& Saunders 1992, Rutledge et al. 2000) 
if we could
compute our $LR$ values using a good estimate of $q(f)$, but, as
mentioned above, the small number of sources here leads us to take 
$q(f)$ to be a constant, thereby leaving $LR$ defined only up to a
multiplicative factor, in which case we cannot employ the formalism of 
Sutherland \& Saunders (1992) or Rutledge et al. (2000)
for quantifying the reliability of associations.

\begin{figure}
\epsfig{file=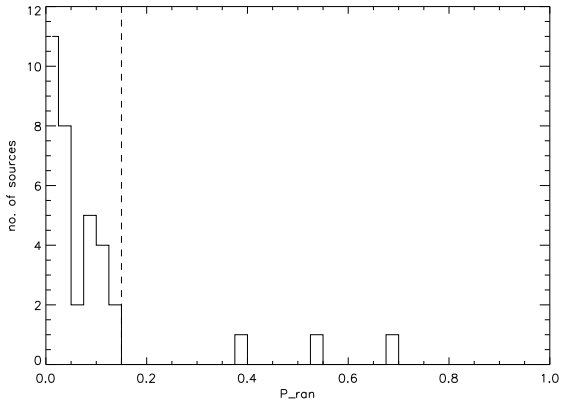,width=8cm}
\caption{The $P_{\rm ran}$ histogram for the associations made between
the 35 sources in Table 8 of Paper I and the R band catalogue of 
Gardner et al. (1999), supplemented by the ESO EIS K or  I band catalogue
for those sources falling into the masked region of the Gardner et
al. image. We define good associations to be those with
$P_{\rm ran}< 0.15$, to the left of the dashed line.}
\label{fig:pran_hist}
\end{figure}

We sought associations for our {\em ISO} sources with objects in the
following set of catalogues: (i) the R band AAT prime focus catalogue
of Verma et al. (2002); (ii) the $I_{814}$--band STScI HST
catalogue\footnote{available from the STScI HDFS WWW site at 
{\tt www.stsci.edu/ftp/science/hdfsouth/catalogs.html}}; (iii)
the I band EIS catalogue of da Costa et al. (1998); (iv) the K band
EIS catalogue of da Costa et al. (1998); and (v) the R band CTIO BTC
catalogue (version 2) of Gardner et al. (1999), available from 
\verb+http://hires.gsfc.nasa.gov/~research/hdfs-btc/+.
Inspection of the results of this process revealed that almost all
of the {\em ISO} sources were associated with the same star or galaxy in
each case, indicating the robustness of this method of association.
Moreover, the associations made using the different catalogues in
their different bands yielded very similar $P_{\rm ran}$ values 
for the associations between given pairs of source and object, giving
us confidence that it is a sound statistic to use.

\subsection{Results}

As mentioned above, very similar $P_{\rm ran}$ results are obtained
for our sources when the likelihood ratio procedure is run against all
five optical and near-IR catalogues listed in the previous subsection.
We take our principal association to be with the R band CTIO
BTC catalogue of Gardner et al. (1999), simply because it includes the
largest number of the objects associated with our sources. 
In Figure \ref{fig:pran_hist} we plot the histogram of $P_{\rm ran}$
values resulting from the association of the {\em ISO} sources in Table
8 of Paper I with objects in the CTIO catalogue. A few of
our sources are clearly associated with objects in regions of the
image masked out by Gardner et al. (1999) when they constructed their
SExtractor (Bertin \& Arnouts 1996) catalogue from it: for these cases
we use the $P_{\rm ran}$ value for associated EIS K band or I band
object instead. This histogram is similar to what we should expect:
most of the {\em ISO} sources have been associated  with objects with low 
$P_{\rm ran}$ values, indicating a high probability that this is a
correct identification, while there is a tail to high $P_{\rm ran}$,
due to sources that are either spurious or have optical counterparts
too faint for detection in the  CTIO image, which has a limiting magnitude
of $R \sim 23$. The $P_{\rm ran}$
value to be taken as the threshold below which associations are defined to
be good is somewhat arbitrary when the number of sources is low, as
here: with a much larger number it is possible to see the point where
the number of sources begins significantly to exceed the flat tail due
to sources which are spurious or have no counterpart in the object
catalogue, but that is clearly not
possible in Figure \ref{fig:pran_hist}. 
The form of that histogram suggests that $P_{\rm ran}=0.15$
might be an appropriate threshold.
Note that Mann et al. (1997) took their threshold
for reasonable identifications of {\em ISO--HDF--N\/} sources to be 0.35, so
the much tighter bunching to lower $P_{\rm ran}$ values here indicates
once more that the {\em ISO} data presented here are of a far higher quality
than those used in the initial {\em ISO--HDF--N\/} analysis. In particular
we have benefitted from the knowledge of the {\em ISO}--CAM image distortion
which was uncharacterized at the time that Mann et al. (1997)
performed their identification of the optical counterparts of the {\em ISO--HDF--N\/}
sources.

\begin{figure*}
\caption{Postage stamp images for the 32 sources from Paper I 
with reliable associations. {\em ISO} contours from the 6.7$\mu$m (dashed) and
15$\mu$m (solid) signal/noise map are plotted over an optical/near-infrared image, centred
on the object associated with the {\em ISO} source: the title of each image gives 
the name of the {\em ISO} source and the optical/NIR image used for the background. 
{\bf High resolution version of this figure available from astro.ic.ac.uk/hdfs.}}
\label{fig:id_post}
\end{figure*}

\begin{figure*}
\contcaption{Postage stamp images for the 32 sources from Paper I 
with reliable associations. {\em ISO} contours 
($1, 2, 3 \ldots 9, 10, 20, \ldots, 90, 100, 200$) from the 6.7$\mu$m (dashed) and
15$\mu$m (solid) signal/noise map are plotted over an optical/near-infrared image, centred
on the object associated with the {\em ISO} source: the title of each image gives 
the name of the {\em ISO} source and the optical/NIR image used for the background.
{\bf High resolution version of this figure available from astro.ic.ac.uk/hdfs.}}
\end{figure*}

\begin{figure*}
\contcaption{Postage stamp images for the 32 sources from Paper I 
with reliable associations. {\em ISO} contours 
($1, 2, 3 \ldots 9, 10, 20, \ldots, 90, 100, 200$) from the 6.7$\mu$m (dashed) and
15$\mu$m (solid) signal/noise map are plotted over an optical/near-infrared image, centred
on the object associated with the {\em ISO} source: the title of each image gives 
the name of the {\em ISO} source and the optical/NIR image used for the background.
{\bf High resolution version of this figure available from astro.ic.ac.uk/hdfs.}}
\end{figure*}

\begin{figure*}
\contcaption{Postage stamp images for the 32 sources from Paper I 
with reliable associations. {\em ISO} contours 
($1, 2, 3 \ldots 9, 10, 20, \ldots, 90, 100, 200$) from the 6.7$\mu$m (dashed) and
15$\mu$m (solid) signal/noise map are plotted over an optical/near-infrared image, centred
on the object associated with the {\em ISO} source: the title of each image gives 
the name of the {\em ISO} source and the optical/NIR image used for the background.
{\bf High resolution version of this figure available from astro.ic.ac.uk/hdfs.}}
\end{figure*}

\begin{figure*}
\contcaption{Postage stamp images for the 32 sources from Paper I 
with reliable associations. {\em ISO} contours 
($1, 2, 3 \ldots 9, 10, 20, \ldots, 90, 100, 200$) from the 6.7$\mu$m (dashed) and
15$\mu$m (solid) signal/noise map are plotted over an optical/near-infrared image, centred
on the object associated with the {\em ISO} source: the title of each image gives 
the name of the {\em ISO} source and the optical/NIR image used for the background.
{\bf High resolution version of this figure available from astro.ic.ac.uk/hdfs.}}
\end{figure*}

\begin{figure*}
\contcaption{Postage stamp images for the 32 sources from Paper I 
with reliable associations. {\em ISO} contours ($1, 2, 3 \ldots 9, 10, 20, \ldots, 90, 100, 200$) 
from the 6.7$\mu$m (dotted) and
15$\mu$m (solid) signal/noise map are plotted over an optical/near-infrared image, centred
on the object associated with the {\em ISO} source: the title of each image gives 
the name of the {\em ISO} source and the optical/NIR image used for the background.
{\bf High resolution version of this figure available from astro.ic.ac.uk/hdfs.}}
\end{figure*}

In Figure \ref{fig:id_post} we show postage stamp images for the 32
sources in Paper I with reliable associations, 
overlaying {\em ISO} 6.7 and 15 $\mu$m contours onto optical images, with the
associated object lying at $(0,0)$. Table 
\ref{tab:id_props} summarizes the properties of these
objects, while the next  subsection presents more details of the associations.
As mentioned previously, principal associations were made with the GSFC R band 
catalogue of Gardner et al. (1999), so the optical magnitudes tabulated are 
from their catalogue, with one exception (ISOHDFS~J223237-603235), which is 
omitted from their catalogue, causing us to use the EIS optical catalogue of 
da Costa et al. (1998). The photometric data in the GSFC catalogue are calibrated 
using standards measured in Johnson UBV and Cousins RI, so we quote all optical 
magnitudes in that system. For the one case of ISOHDFS~J223237-603235, this involves 
converting the UBVRI magnitudes back to the Johnson-Cousins system from the AB system 
in which da Costa et al. (1998) present their photometry, using the conversions given 
in their paper, namely $U=U_{\rm AB}-0.82$, $B=B_{\rm AB}+0.06$, $V=V_{\rm AB}$, 
$R=R_{\rm AB}-0.17$ and $I=I_{\rm AB}-0.42$. For their near-infrared data, da Costa 
et al. (1998) performed photometric calibration in JHK using Persson et al. (1998) 
standards and then converted them to AB magnitudes: again we have reversed this 
conversion, using the relations quoted by da Costa et al. (1998), namely 
$J=J_{\rm AB}-0.89$, $H=H_{\rm AB}-1.38$ and $K=K_{\rm AB}-1.86$. In Table 
\ref{tab:id_props} we present spectroscopic redshifts from Glazebrook et al. (2002) 
where available, and photometric redshifts where not. Photometric redshifts were 
estimated by three of us, using independent methods: M. Rowan--Robinson (MRR) used
(Rowan--Robinson 2001) SEDs based on those of Yoshii \& Takahara (1988), A. Franceschini (AF) 
used the PEGASE
models of Fioc \& Rocca--Volmerange (1997) and R.G. Mann (RGM) used the GRASIL (Silva et al. 1998)
model fits to SEDs of
a range of known galaxies (Arp220, M100, M51, M82, NGC6090, NGC6946) presented by Silva et al. 
(1998). Results for individual galaxies are
discussed below, but, in most cases, these independent methods yielded photometric redshift
estimates that agreed with each other, and with those computed by Stephen Gwyn (Gwyn 1999) 
and the SUNY group\footnote{See {\tt www.ess.sunysb.edu/astro/hdfs/home.html}}, to 
$\delta z \sim 0.1$, which is therefore the accuracy we claim 
for the adopted photometric redshifts listed in parentheses in Table \ref{tab:id_props}.

\begin{table*}

\begin{minipage}{170mm} 
\caption{Properties of the objects associated with the
{\em ISO} sources in the Hubble Deep Field South.}
\label{tab:id_props}

\begin{tabular}{lcccccccccccc}

{\em ISO} Source & RA\footnote{All RAs prefixed by 22$^{\rm h}$.} & Dec\footnote{All Decs prefixed by -60$^{\circ}$.} & $P_{\rm ran}$ & U & B & V & R & I & J & H & K & z\\ 

  \hline

 J223237-603256 &   32 37.4 &  32 57.6 & 0.001 & 20.53 & 19.54 & 17.93 & 16.77 & 15.49 & & & & (0.00)\\ 
 J223237-603235\footnote{The object associated with this source is not included in the GSFC catalogue, so the tabulated photometric data come from the EIS catalogue of da Costa et al. (1998)} &  32 37.9 & 32 33.5 & 0.001 & 20.38 & 19.33 & 17.91 & 17.12 & 16.48 & & & & (0.00)\\
 J223240-603141 &   32 40.6 &  31 43.6 & 0.095 & 22.88 & 22.82 & 22.09 & 21.19 & 20.46 & & & & (0.45)\\ 
J223243-603242 & 32 43.0 &  32 42.4 & 0.047 & 22.91 & 22.92 & 21.64 & 20.65 & 19.85 &18.57 &17.75 &16.89 &(0.50) \\
 J223243-603441 &  32 43.5 &  34 42.3 & 0.021 & 21.60 & 21.71 & 20.56 & 19.63 & 18.94 &17.71 &16.83 &16.14 &(0.50) \\  
 J223243-603351 &  32 43.5 &  33 51.6 & 0.028 & 19.98 & 20.35 & 20.34 & 19.92 & 19.50 &18.96 &18.13 &17.81 &0.0918 \\  
 J223244-603455 &  32 44.1 &  34 57.2 & 0.031 & 21.50 & 21.71 & 20.68 & 19.91 & 19.28 &18.15 &17.34 &16.70 & (0.35)\\  
 J223244-603110 &  32 44.3 &  31 11.4 & 0.066 &22.08 &22.13 &21.43 &20.74 &20.23 & 19.70 & 19.03 & 18.26 & 0.25 \\
 J223245-603418 &  32 45.6 &  34 18.9 & 0.078 & 23.55 & 23.20 & 22.02 & 21.19 & 20.45 &18.94 & 17.87 &17.08 &0.4606\\  
J223245-603226 & 32 45.8 &  32 26.3 & 0.080 &23.26 &23.15 &22.28 &21.29 &20.59 &19.33  & 18.39 & 17.66 & 0.59\\  
 J223247-603335 & 32 47.7 & 33 35.9 & 0.030 & 21.99 & 22.06 & 21.06 & 20.03 & 19.24 &17.97 &17.01 &16.31 & 0.5803\\  
 J223250-603359 & 32 50.5 & 34 00.8 & 0.003 & 21.44 & 20.32 & 18.78 & 17.78 & 16.75 &15.53 &14.83 &14.63 & (0.00)\\  
 J223251-603335 & 32 51.5 &  33 37.7 & (0.102)& 23.87 & 24.36 & 23.44 & 22.50 & 21.89 &20.72 &19.58 &18.92 &(0.56)\\
 J223252-603327 & 32 53.0 &  33 28.6 & (0.073) & 24.91 & 24.80 & 24.10 & 23.33 & 22.37 &20.60 &19.57 &18.96& 1.27\\  
 J223254-603129 & 32 54.8 & 31 31.1 & 0.123 & 23.22 & 23.78 & 22.53 & 21.48 & 20.84 &20.47 &19.45 &18.81 & (0.20)\\  
 J223254-603143 & 32 54.9 & 31 44.1 & 0.009 & 23.39 & 21.92 & 20.36 & 19.22 & 18.02 &16.62 &15.87 &15.65 & (0.00)\\  
J223254-603115\footnote{The position of the object associated with this source in the GSFC catalogue suggests that the optical magnitudes quoted here may be too bright, through the inclusion of a close, faint companion, which  the EIS near-infrared catalogue marks as a separate object, but which the GSFC catalogue does not.} & 32 54.8 &  31 14.6 & 0.044 & 23.52 & 23.22 & 21.73 & 20.51 & 19.67 &20.23 &19.26 &18.57 & 0.5111\\  
 J223257-603305 & 32 57.5 & 33 06.0 & 0.047 & 21.87 & 22.09 & 21.44 & 20.66 & 20.01 &19.00 &18.09 &17.45 & 0.5823\\
 J223259-603118 & (32 59.5 & 31 19.1) & ($<0.001$) & & & & & & 12.68 & 12.22 & 12.17 & (0.00)\\
 J223302-603213 & 33 02.7 & 32 13.8 & 0.006 & 22.76 & 21.52  & 19.94 & 18.78 & 17.46 &16.01 &15.38 &15.15 & 0.00\\
 J223302-603323 & 33 02.8 & 33 22.4 & 0.049 &22.74 &22.82 &21.68 &20.70 & 19.95 & 18.62 &17.70 & 16.90 & (0.60)\\
 J223303-603230 & (33 03.1 & 32 30.8) & (0.001) & & & & & & 14.08 & 13.35 & 13.18 & 0.00\\
 J223303-603336 & 33 03.6 &  33 41.7 & 0.131 &22.32 &22.34 &21.14 &20.40 &19.78 &18.71  &17.84 &17.20 & (0.35)\\ 
 J223306-603436 & 33 05.8 & 34 37.2 & 0.095 &24.37 &23.93 &22.55 &21.34 &20.44 & 19.00 & 17.89 &17.08  & (0.60)\\ 
 J223306-603349 & 33 06.0 & 33 50.3 & 0.002 &18.78 &18.79 &17.80 &17.29 &16.67 &15.79  &15.07  & 14.50 & 0.1733\\  
 J223306-603450 & 33 07.0 & 34 51.7 & (0.094) &25.29 &25.31 &24.79 &23.46 &22.45 & 21.08 & 20.01 & 19.07 & (0.75)\\
 J223307-603248 & 33 07.6 & 32 50.3 & 0.034 &22.20 &22.20 &21.18 &20.21 &19.54 &18.37  & 17.43 & 16.73 & 0.513\\
 J223308-603314 & 33 08.2 & 33 21.6 & 0.004 &20.05 &18.98 &17.87 &17.21 &16.66 &15.86  &15.19  &15.11  & (0.50)\\
 J223312-603416 & (33 12.1 & 34 16.7) & (0.128) & & & & & & 21.57 & 20.38 & 19.34 & (1.30)\\
 J223312-603350 & 33 12.6 & 33 50.5 & 0.103 &23.20 &23.42 & 22.40&21.41 &20.34 &19.22  &18.60  & 18.15 & (0.50)\\
 J223314-603203 & (33 14.3 & 32 06.0) & (0.111) & & & & & &  20.77 & 19.68 & 18.98 & (0.20)\\
 J223315-603224 & (33 15.8 & 32 34.0) & ($<0.001$) & & & & & & 11.58 & 10.86 & 10.67 & 0.00\\ 

\end{tabular}

\medskip

Notes to Table \protect\ref{tab:id_props}: The positions and $P_{\rm ran}$ values tabulated refer to the associated object in the GSFC catalogue, with the exception of the 8 sources which have $P_{\rm ran}$ values in parentheses, which denote that they refer to the association with an object in the EIS near-infrared catalogue. For five of these cases (those with positions in parentheses) this is because the source is missing from the GSFC catalogue, while the remaining have significantly lower $P_{\rm ran}$ values resulting from their likelihood ratio association with the EIS K band catalogue than with the GSFC R band catalogue, as a result of red $R-K$ colours: these three objects are the first, second and ninth reddest of the 24 associated objects for which $R-K$ has been measured.

\end{minipage}
\end{table*}

\subsection{Notes on individual sources}\label{sect:id_notes}

\begin{enumerate}

\item {\bf ISOHDFS~J223237-603235:} The association here is with an
I=16 star, ~2 arcsec from the {\em ISO} source position and yielding 
$P_{\rm ran}=0.001$ from the EIS optical
catalogue (the star is masked out of the CTIO catalogue).

\item {\bf ISOHDFS~J223237-603256:} As shown by the contours in
Figure~\ref{fig:id_post}, this 6.7 $\mu$m source is very safely
($P_{\rm ran}=0.001$) associated with a bright (17th magnitude in R) 
object clearly identified as a star by SExtractor in the AAT R and EIS
I images. 

\item {\bf ISOHDFS~J223240-603141:}  This 15 $\mu$m source is
associated fairly securely ($P_{\rm ran}=0.095$) with an I=20 galaxy,
located ~3 arcsec from the {\em ISO} position. MRR estimates $z_{\rm phot}=2.02$
for this galaxy, on the basis of its UBVRI magnitudes, while both AF and RGM
estimate 0.45, so it is this latter figure we adopt: as shown in Figure~\ref{fig:seds}, 
with this assumed redshift, the optical and {\em ISO} data for this
galaxy are well fitted by the GRASIL model for NGC6090.

\item {\bf ISOHDFS~J223243-603242:} This is an I=20 galaxy,
detected significantly ($>4 \sigma$) in the co-added maps at both 6.7
and 15 $\mu$m. This association yields $P_{\rm ran}=0.047$, and
MRR and AF both estimate a photometric redshift of $z_{\rm phot}=0.5$
from the optical/NIR data
for this galaxy, while RGM obtains a slightly lower value of 0.45: we adopt
0.5, and show in Figure~\ref{fig:seds} that with this redshift, the U band to 
6.7$\mu$m data are in good agreement with the GRASIL model for the starburst
galaxies (Arp220, M82, NGC6090).

\item {\bf ISOHDFS~J223243-603351:} Another very significant
detection at both 6.7 and 15 $\mu$m, this source is associated
($P_{\rm ran}=0.028$) with a
19th magnitude galaxy, with detections at both 4.9 and 8.6 GHz (with
fluxes of 0.163 and 0.111 mJy respectively: A. Hopkins, {\em priv. comm.\/}). 
Its spectrum exhibits one
broad line, and yields a redshift of $z=0.0918$, so this appears to be
a low-redshift AGN:  RGM, AF, MRR estimated $z_{\rm phot}=0.00, 0.15, 0.12$, 
respectively.

\item {\bf ISOHDFS~J223243-603441:} This source, detected very
significantly in both {\em ISO} bands, is associated ($P_{\rm ran}=0.021$)
with the brighter (19th magnitude in I) of a close pair of galaxies,
for which RGM, AF, MRR estimate $z_{\rm phot}=0.40, 0.50, 0.59$, respectively,
on the basis of UBVRIJHK photometry.
We adopt $z=0.5$ and show in Figure~\ref{fig:seds} that this gives this galaxy
an SED more like the cirrus galaxy M51 than the starbursts.

\item {\bf ISOHDFS~J223244-603110:} This source, solidly detected
at 15 $\mu$m, is located 3 arcsec from an I=20 galaxy, with which we
associate it securely ($P_{\rm ran}=0.066$). The UBVRIJHK photometry
for this galaxy yields a fairly wide spread of photometric
redshift estimates, with RGM, AF and MRR obtaining 0.10, 0.25 and 0.32,
respectively. We adopt the middle value, using which the galaxy's SED
is seen to be similar to that of the GRASIL fit to the starburst NGC6090,
and note that this is one of our more uncertain redshift estimates.

\item {\bf ISOHDFS~J223244-603455:} This 15$\mu$m source is
associated with an I=19 galaxy 2 arcsec from the {\em ISO} position, with a
$P_{\rm ran}$ value of 0.031, and UBVRIJHK magnitudes yielding
photometric redshift estimates of
0.32, 0.30 and 0.45 from MRR, RGM and AF, respectively. A value of
$z_{\rm phot}=0.35$ is adopted, with which the SED of this galaxy 
matches that of the GRASIL model for the normal spiral M100, as shown in 
Figure~\ref{fig:seds}.

\item {\bf ISOHDFS~J223245-603226:} This 15 $\mu$m source is
associated with one of a pair of close (interacting?) galaxies, with
I=20 and $P_{\rm ran}=0.080$, for which RGM, AF and MRR estimate photometric
redshifts of 0.50, 0.50 and 0.51. Rigopoulou et al. (2000) measured a
spectroscopic redshift of $z=0.59$, and, using that value, this galaxy's
SED is a reasonable fit to the GRASIL model SED for the starburst NGC6090.

\item {\bf ISOHDFS~J223245-603418:} This is an I=20 galaxy, detected at
both 6.7 and 15 $\mu$m, yielding $P_{\rm ran}=0.076$, with a spectrum
showing H$\beta$ and O{\sc iii} (4959+5007) lines, from which a
redshift of $z=0.4606$ was determined:  RGM, AF, MRR estimated 
$z_{\rm phot}=0.55, 0.6, 0.95$, respectively, for this source.
This is also a radio source,
detected at 1.4, 2.5 and 4.9 GHz, with fluxes of 0.200, 0.149 and
0.127 mJy, respectively (A. Hopkins, {\em priv. comm.\/}). As shown by 
Figure~\ref{fig:seds}, the SED of
this galaxy is in good agreement with the GRASIL model for Arp220 over
six decades in wavelength.

\item {\bf ISOHDFS~J223247-603335:} Detected at both 6.7 and 15
$\mu$m, we associate ($P_{\rm ran}=0.029$) this source with an I=19
galaxy, with a spectrum displaying a number of emission lines and
yielding a redshift of $z=0.5803$: RGM AF, MRR, Gwyn and SUNY group
estimated photometric redshifts of 0.50, 0.60, 0.52, 0.56, 0.66, respectively,
for this galaxy, whose SED is similar to that of the GRASIL model for the
spiral galaxy M51.

\item {\bf ISOHDFS~J223250-603359:} This 6.7 $\mu$m is associated
very securely ($P_{\rm ran}=0.003$) with an I=17 M2V star displaying clear
diffraction spikes in its WFPC2 image.

\item {\bf ISOHDFS~J223251-603335:} This is identified with an I=22 galaxy 
3 arcsec away, yielding $P_{\rm ran}=0.102$ for association with the
EIS K band catalogue. RGM, AF, MRR, Gwyn and the SUNY group estimate
$z_{\rm ph}$= 0.50, 0.70, 0.95, 0.56, 0.57, respectively, for this galaxy, whose
SED (with an adopted redshift of 0.7) shows the rise through the infrared 
characteristic of a starburst galaxy.

\item {\bf ISOHDFS~J223252-603327:} This is a radio source (0.109
mJy at 1.4 GHz: A. Hopkins, {\em priv. comm.\/}), located at the centre of 
quite an extended region of 
emission in both {\em ISO} bands. The source is associated ($P_{\rm
ran}=0.073$)  with the EIS K band
catalogue. SExtractor stellarity indices exist for this object in U,
B, V, R, I, J, H and K bands from the EIS catalogue, and vary from
0.48 in U to 0.97, indicating that the image is getting more point-like
at longer wavelengths. This, together with the radio detection,
suggests that this may be an obscured AGN, although Rigpoulou et al.
(2000) do not report any strong AGN features in their spectrum of this
object. That yielded $z=1.27$, in excellent agreement with the 
photometric redshifts of Gwyn and the SUNY group (1.28 and 1.27, respectively), 
while RGM, AF and MRR obtained more widely--varying values of 0.50, 0.90 and 1.46
for this galaxy.

\item {\bf ISOHDFS~J223254-603115:} This source, detected at both
6.7 and 15 $\mu$m, is associated with the brighter of a pair of very close
(interacting?) galaxies separated by ~2 arcsec. In the ESO EIS
near-infrared catalogue of da Costa et al. (1998), the two galaxies
have (J,H,K) magnitudes of (21.12, 20.64, 20.43) and (21.63, 21.18,
21.03), while the brighter galaxy has I=19.71 in the GSFC CTIO
catalogue, yielding $P_{\rm ran}=0.044$, although it appears that the
optical magnitudes include a significant contribution from the second
galaxy, and, as a result the optical and near--infrared photometry in 
Figure~\ref{fig:seds} do not appear to be consistent. Fortunately, a 
spectroscopic redshift of 
$z=0.5111$ was measured for this galaxy by Glazebrook et al. (2002),
so it is not necessary to try to estimate a photometric redshift from this
inconsistent set of magnitudes. Assuming that the near--infrared photometry
is correct, the SED obtained using $z=0.5111$ is very similar to that for the
GRASIL model for the starburst galaxy NGC6090.

\item {\bf ISOHDFS~J223254-603129:} This is an intriguing
case. The peaks of 6.7 and 15$\mu$m emission are very close to each other, but 4-5
arcsec away (in directions 120$^\circ$ apart) from both an I=21 galaxy
(yielding $P_{\rm ran}=0.162$) and a 0.143 mJy 1.4 GHz radio
source, which appear not to be associated. It is possible that the
{\em ISO} source is not associated with that galaxy either (but, instead,
with an object too faint for these optical/near--infrared survey data that
may or may not be the source of the radio emission), but we assume here 
that it is. RGM, MRR and AF estimated redshifts of 0.15, 0.48 and 0.50 for this
galaxy, and, adopting $z_{\rm phot}=0.2$ we see, from Figure~\ref{fig:seds}, that
we obtain an SED similar to that of the GRASIL model for Arp220.

\item {\bf ISOHDFS~J223254-603143:} This 6.7$\mu$m source is
associated securely ($P_{\rm ran}=0.009$) with an I=18 star, which
shows clear diffraction spikes in HST imaging data.

\item {\bf ISOHDFS~J223256-603059:}  This 6.7 $\mu$m source has no
reasonable association: its best association in the GSFC CTIO
catalogue is with an I=22.5 galaxy 5 arcsec away, yielding 
$P_{\rm ran}=0.398$. The source is located very close to the very
bright source ISOHDFS~J223259-603118, and may be an artifact
produced by inaccuracies in the application of our background
subtraction procedure so close to this, the second brightest 6.7
$\mu$m source in our catalogue.

\item {\bf ISOHDFS~J223256-603513:} There is no reasonable
association for this 15 $\mu$m source in the GSFC CTIO catalogue, which is the
only one covering this area: the best association is with an I=22
galaxy 7 arcsec away, which yields $P_{\rm ran}=0.697$.

\item {\bf ISOHDFS~J223257-603305:} This 15 $\mu$m source is
associated ($P_{\rm ran}=0.047$) with an I=20 galaxy for which
Glazebrook et al. (2002) determined a redshift of $z=0.5823$ from a
spectrum exhibiting a number of narrow emission lines, which, together
with the relative levels of the 15$\mu$m detection and 6.7$\mu$m upper
limit, suggest that this is a starburst galaxy, although its SED is not
a particularly good match to any of the GRASIL models.  RGM, AF, MRR had 
estimated $z_{\rm phot}=0.40, 0.60, 0.62$, respectively, for this source.

\item {\bf ISOHDFS~J223259-603118:} This strong 6.7$\mu$m source is
a bright star, detected also at 15 $\mu$m but masked out of most of
the SExtractor catalogues created from the optical/near--IR surveys
under discussion here. It appears in the EIS survey as a K=14 G2III
star,
yielding $P_{\rm ran} < 0.001$.

\item {\bf ISOHDFS~J223302-603137:} Like ISOHDFS~J223256-603059,
this 6.7 $\mu$m source has no reasonable identification: the best
candidate in the GSFC CTIO catalogue is a faint (I=25) galaxy lying
almost 5 arcsec away, and yielding $P_{\rm ran}=0.539$. This source
lies close to the bright star ISOHDFS~J223259-603118, and, like the
other unidentified 6.7$\mu$m source, ISOHDFS~J223256-603059, may be
an artifact resulting from background subtraction errors.
 
\item {\bf ISOHDFS~J223302-603213:} The object associated with this
source is bright (I=17) and less than 1 arcsec from the {\em ISO} position,
yielding a $P_{\rm ran}$ value of 0.006. It is classed as stellar by 
SExtractor, and classified as an M3V star on the basis of the spectrum 
taken by Glazebrook et al. (2002).

\item {\bf ISOHDFS~J223303-603230:} The association for this
source, detected significantly at both 6.7 and 15 $\mu$m, is a bright
object, spectroscopically confirmed to be an M1V star using the
Glazebrook et al. (2002) spectrum. 
It is masked out of the GSFC CTIO catalogue, but has K=15
in the EIS near-infrared catalogue, yielding $P_{\rm ran}=0.001$.

\item {\bf ISOHDFS~J223302-603323:} This source, detected in both
bands, is associated ($P_{\rm ran}=0.049$) with an I=20 galaxy for
which RGM, AF, MRR, Gwyn and the SUNY group estimate photometric redshifts of 
0.40, 0.60, 0.66, 0.474 and 0.400, respectively. With an adopted 
$z_{\rm phot}=0.60$, we obtain an SED similar to that of the GRASIL models
for normal spirals like M100, NGC6946 or M51.

\item {\bf ISOHDFS~J223303-603336:} The I=20 galaxy associated with
this source is 6 arcsec from the {\em ISO} position, which is the cause of
its relatively poor $P_{\rm ran}$ value of 0.131, but the {\em ISO} source
position could be shifted due to the close proximity of the bright
source ISOHDFS~J223306-603349 (the brightest in our 15$\mu$m
catalogue): once again, our background subtraction method could be
leaving artifacts close to this bright source. RGM, AF, MRR, Gwyn
and the SUNY group estimate photometric redshifts of 0.30, 0.35, 0.35, 
0.419 and 0.440, respectively, for this galaxy,  and with an adopted
redshift of 0.35 we obtain an SED which fits the GRASIL starburst models
well from the U band to 6.7$\mu$m: we assume that the absence of a detection 
at 15$\mu$m is due to problems with the subtraction of the side--lobes of 
ISOHDFS~J223306-603349.

\item {\bf ISOHDFS~J223306-603349:}  This I=16 spiral galaxy, with a
spectroscopic redshift determined by Glazebrook et al. (2002) to be
$z=0.1733$, is the brightest 15 $\mu$m source in our catalogue, and
has $P_{\rm ran}=0.002$. It is detected in the radio at 0.533 and
0.300 mJy at 1.4 and 2.5 GHz, respectively (A. Hopkins, {\em priv. comm.\/}), 
and, from Figure~\ref{fig:seds},
we see that  this galaxy has the SED of a normal spiral galaxy, rather than
a starburst.  RGM, AF, MRR estimated $z_{\rm phot}=0.15, 0.25, 0.15$, respectively,
for this galaxy.

\item {\bf ISOHDFS~J223306-603436:} This I=20 galaxy yields a
$P_{\rm ran}$ value of 0.095, and is detected in both {\em ISO} bands. MRR,
RGM and AF estimate photometric redshifts of 0.35, 0.60 and 1.00 for this
galaxy, and an adopted $z_{\rm phot}=0.60$ gives an SED very similar to that
of the GRASIL models for the starburst galaxies M82 and NGC6090.

\item {\bf ISOHDFS~J223306-603450:} The identification of this
source is complicated. Its {\em ISO} source position is less than 1 arcsec
from a faint galaxy (I=22), which is found with K=21 in the EIS
catalogue. The $P_{\rm ran}$ value resulting from the association of
this object in the GSFC CTIO catalogue is 0.307, but it is a much
better 0.094 when computed in the K band, indicating that this galaxy
is very red in $R-K$. RGM, MRR and AF estimate redshifts for it of 0.65,
1.05 and 0.90, and with an adopted $z_{\rm phot}=0.75$ we obtain an SED
similar to the GRASIL model of Arp220.

\item {\bf ISOHDFS~J223307-603248:} This source, detected in both
bands, is securely ($P_{\rm ran}=0.034$) associated with an I=20 galaxy 
for which Glazebrook et al. (2002) have determined a spectroscopic redshift of
$z=0.513$, yielding an SED similar to the GRASIL model for the
starburst NGC6090.  RGM, AF, MRR estimated $z_{\rm phot}=0.40, 0.50, 0.55$, respectively,
for this galaxy.

\item {\bf ISOHDFS~J223308-603314:} The likelihood ratio procedure
associates (with $P_{\rm ran}=0.034$) this source with an I=17 K4V
star,
although the {\em ISO} contours in both bands are centred closer to an 
I=20 galaxy 5 arcsec away. AF, MRR and RGM estimate redshifts of 0.05,
0.10 and 0.50, and we choose the last of these, as that gives the best fit to one
of the GRASIL SEDs, that for the NGC6090 model, but highlight that this is one
of our most uncertain redshift estimates.

\item {\bf ISOHDFS~J223312-603350:} Detected significantly in both
bands, this source is associated (with $P_{\rm ran}=0.103$) with an 
I=20 galaxy 2 arcsec from the 6.7$\mu$m source position  and 3 arcsec
from that at 15$\mu$m. Glazebrook et al. (2002) took a spectrum at
this sky position, but it yielded no features capable of determining
the galaxy's redshift. RGM, AF and MRR determined photometric redshifts
of 0.45, 0.70 and 0.32, respectively, but none of these yields an SED
in particularly good agreement with the GRASIL starburst models: we adopt
$z_{\rm phot}=0.50$, as that gives, perhaps, the best agreement  (with the
Arp220 SED), but note that this is highly uncertain.

\item {\bf ISOHDFS~J223312-603416:} This {\em ISO} source position lies
1 arcsec from a faint (K=21) EIS galaxy, which is not present in the
GSFC CTIO catalogue. This near-infrared association yields $P_{\rm
ran}=0.128$, and MRR, AF and RGM estimate redshifts of 1.24, 1.30
and 2.05, respectively, although this is clearly very uncertain, since it
is determined from JHK photometry alone. The relative fluxes in the
two {\em ISO} and three NIR bands suggests that this is a starburst galaxy, but
none of the estimated redshifts listed above results in a good fit to any
of the GRASIL SED models; for definiteness we assume $z_{\rm phot}=1.3$,
but stress that this is not at all well--constrained.

\item {\bf ISOHDFS~J223314-603203:} This is another example of
a background subtraction artifact complicating the identification of a
source. This source, detected significantly in both bands, 
lies close to a very bright star (the brightest 6.7 $\mu$m source in
our catalogue, and the second brightest at 15$\mu$m) and has no
reliable association. The region is masked out of the GSFC CTIO
catalogue, while the EIS near--infrared catalogue produces an
association with $P_{\rm ran}=0.111$ with a K=21 galaxy ~3 arcsec
from the {\em ISO} source position, but the contours in this region have
probably been disturbed by the background subtraction procedure.
RGM and MRR estimated redshifts of 0.45 and 0.82, respectively, for this
galaxy, on the basis of JHK photometry only, while we note that a
significantly lower redshift of $z_{\rm phot}=0.20$ produces a reasonable
match to the SED of the starburst NGC6090, so we adopt that here: it is
clear, however, that this is one of our most uncertain associations.

\item {\bf ISOHDFS~J223315-603224:} This is the brightest source in
our catalogue at 6.7$\mu$m and the second brightest at 15$\mu$m. It is a
bright (K=13) M2V star, and is 
masked out of the GSFC CTIO catalogue, but its association with the
EIS near-infrared catalogue yields $P_{\rm ran} < 0.001$.

\end{enumerate}

In summary, we have found associations for 32 out of the 35 {\em ISO\/} sources
from Paper I. Of the remaining three, two (ISOHDFS~J223256-603059 and ISOHDFS~J223256-603137)
were detected only at 6.7$\mu$m and lie very close to sources which are very bright in that
band, and which may well have compromised the background subtration procedure in that area,
resulting in a false detection, or a significant shift in the source position. The third,
ISOHDFS~J223256-603513, was detected only at 15$\mu$m, and is a peak in an extended region
of emission stretching from a bright source just below the southern boundary of our source
detection region (see Figure 6 of Paper I). Eight of the 32 identified sources are stars,
leaving a total of 24 galaxies to be considered further.

\begin{figure*}
\hspace{-0.5cm}
\epsfig{file=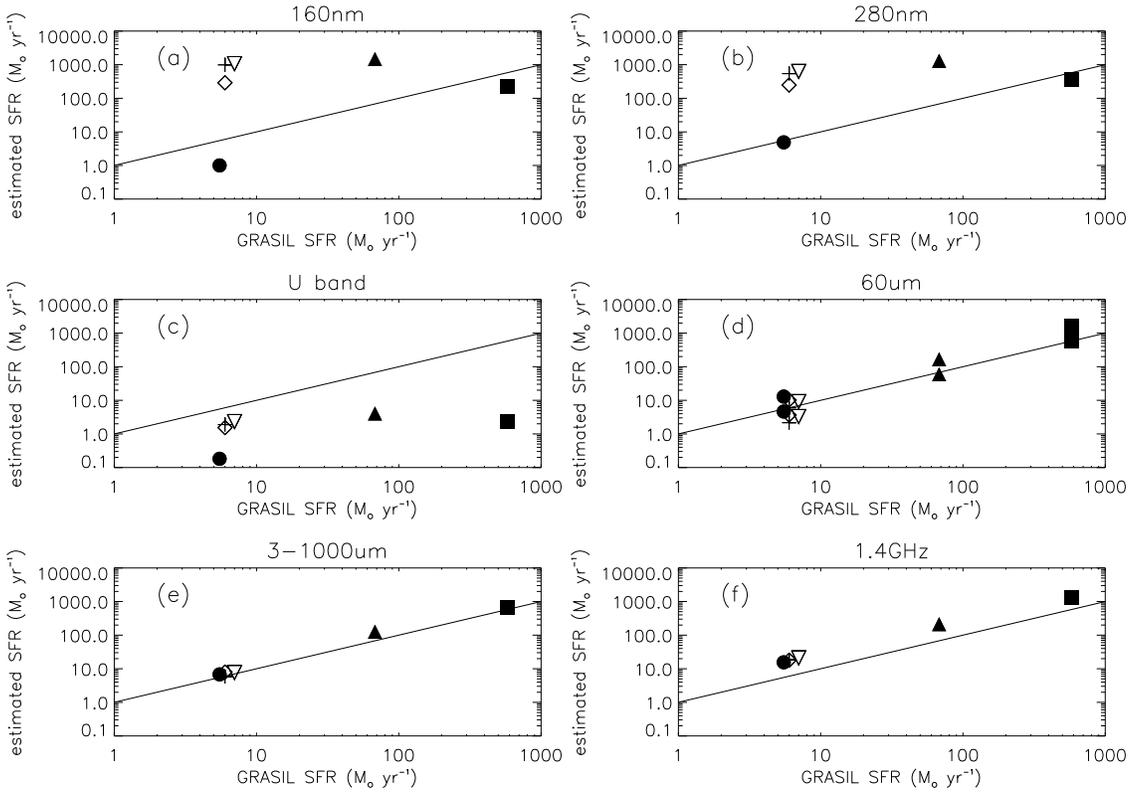,width=15cm, angle=0}
\caption{Comparison of the results of applying to {\em GRASIL\/} model SEDs
six star formation rate estimation methods from eqn.~\ref{eq:scaled_sfr}, namely those based on: (a) 
the 160nm luminosity, from
Devriendt et al. (1999); (b) the 280nm luminosity, from Devriendt et al. (1999); (c) the
U band luminosity, from Cram et al. (1998); (d) the 60$\mu$m luminosity from Cram et al. (1998)
and Rowan--Robinson et al. (1999), denoted by the upper and lower sets of points, respectively;
(e) the 3--1000$\mu$m integrated luminosity, from Devriendt et al. (1999); (f) the 1.4GHz luminosity,
from  Cram et al. (1998). For each method we plot the SFR estimate obtained by its application to the 
{\em GRASIL\/} SED models for M100 (inverted triangle), M51 (diamond), NGC6946 (cross), NGC6090
(filled triangle), M82 (filled circle) and Arp220 (filled square) against the corresponding SFR value
computed within the {\em GRASIL\/} model itself.}
\label{fig:sfr_methods}
\end{figure*}
  
\section{Star formation rate estimators}\label{sect:sfr_methods}

In this Section we discuss and compare the various methods by which we might estimate
the star formation rates of our {\em ISO\/} sources.

\subsection{Methods for estimating star formation rates}

A multitude of methods have been advocated for the estimation
of the star formation rates of galaxies, and, as discussed, for example, by Cram et al. 
(1998) and Granato et al. (2000), 
they agree to varying degrees. All attempt to detect the observational consequences
of the formation of massive stars, and then use a model for the stellar initial mass
function (IMF) to infer the total rate of 
formation of stellar mass. This inevitably introduces some uncertainty into the SFR 
estimates, due to the assumption of a universal IMF for all galaxies and the lack of a
clear choice between competing models for it. More serious is the fact that the formation of 
massive stars in external galaxies, like that in our own Galaxy, is expected to take place
in giant molecular clouds, enshrouded by dust, and, as argued, for example, by Silva et al. (1998)
and Jimenez et al.
(1999), the resultant emission from the star--forming region cannot be well represented by 
a simple model such as a dust--screen in front of a dust--free starburst.

Despite these difficulties, there are several well--used prescriptions for estimating
the SFRs of galaxies on the basis of simple broad--band fluxes, as these are, typically,
all that are available for the large samples of galaxies used in censuses of the cosmic
star formation history. The first of these is to use the U band magnitude, arguing that
the UV emission comes principally from massive, young stars. There are two problems with
this. Firstly, the U band ($\lambda_{\rm eff}=365$nm) is not far enough into the 
ultraviolet for the emission from older
stars to be negligible, but it is possible to correct for this effect, by using 
model spectra to bootstrap from the U band to, say, 250nm, where the emission is dominated
by massive stars. Cram et al. (1998) do this, obtaining the following relationship
between the U band luminosity of a galaxy and its rate of massive ($M \geq 5 M_{\odot}$)
star formation:
\begin{equation}
\frac {SFR (M \geq 5 M_{\odot})}{M_{\odot} \;\; {\rm yr}^{-1}} = \frac {L_{\nu}({\rm U})} {1.5 \times 10^{22} \; \; {\rm W Hz}^{-1}},
\label{eq:sfr1}
\end{equation}
for an assumed $\psi(M) \propto M^{-2.5}$ IMF running from 0.1 to 100 $M_{\odot}$. 
An obvious objection to this is that it fails to account for the effects of dust in the
star--forming galaxies, especially in the case that more massive starbursts are dustier. The
recent models of Devriendt, Guiderdoni \& Sadat (1999) seek to model the spectral energy
distributions (SEDs) of dusty starbursts from the UV to the submillimetre in a 
self--consistent fashion, taking account of the absorption of ultraviolet light from massive
young stars by the dust surrounding them, and its use in heating up the dust, leading to
emission in the far--infrared and submillimetre. Their models yield the following 
correlations between the star formation rate (for a Salpeter 1955 IMF over [0.1,120] $M_{\odot}$) 
and the luminosities at two UV wavelengths:
\begin{equation}
\frac {SFR} {M_{\odot} \; \; {\rm yr}^{-1}}  =  \left[ \frac {\lambda  L_{\lambda}({\rm 
280nm})}{7.7 \times 10^{34} {\rm W}} \right]^{1.62}
\label{eq:sfr2}
\end{equation}
and
\begin{equation}
\frac {SFR} {M_{\odot} \; \; {\rm yr}^{-1}}   =    \left[ \frac {\lambda  L_{\lambda}({\rm 
160nm})}{9.2 \times 10^{34} {\rm W}} \right]^{1.72},
\label{eq:sfr3}
\end{equation}
Devriendt et al. (1999) caution against the over--interpretation of these best--fit 
correlations
(about which their model galaxies scatter quite widely), but the non--linearity of the
relationships they describe between SFR and UV luminosity is qualitatively what would be
expected in a model in which the more massive a starburst is the dustier it is, and suggests
that this effect may be corrected for, albeit crudely.

The bolometric luminosities of luminous starburst galaxies are dominated by thermal 
emission from dust in the far--infrared. In general, this emission
is a combination of extended ({\em ``cirrus''}) emission from dust heated by the ambient 
interstellar radiation field, more localized {\em ``starburst''} emission from the dust 
in regions of massive
star formation heated by the UV flux from O and B stars, and, possibly, {\em ``AGN''} 
emission from the hot dusty torus of an active galactic nucleus. The relative importance 
of these components varies between
galaxies (e.g. Rowan--Robinson \& Crawford 1989), with the cirrus component declining to
higher luminosities, where ultra-luminous infrared galaxies (ULIRGs) are seen 
to be mostly fueled by starburst emission, although some do have a dominant AGN
component (Genzel et al. 1998): such simple modeling must be supplemented by consideration of emission from
PAHs and inclusion of starburst components of different ages 
(Silva et al. 1998, Efstathiou et al. 2000) to fit the detailed SEDs revealed 
(e.g. Acosta--Pulido et al. 1996)
by recent {\em ISO} spectroscopy.  
Various attempts have been made to relate this far--infrared emission to the
amount of ultraviolet light that must be absorbed by dust to produce it, by which route
a galaxy's SFR (or, at least, the rate of formation of stars whose light is obscured by
dust) can be estimated from its far--infrared luminosity. Rowan--Robinson et al. (1997)
summarized previous work on this method in the relationship
\begin{equation}
\frac {SFR} {M_{\odot} \; \; {\rm yr}^{-1}}  =  \frac {\lambda L_{\lambda} ({\rm 60 \mu m})}
{1.5 \times 10^{36} {\rm W}} \cdot \left( \frac{\phi}{\epsilon} \right)
\label{eq:sfr4}
\end{equation}
where $\epsilon \simeq 1$ is the fraction of the ($\nu L_{\nu}$) UV luminosity of the starburst that is
re-emitted in the far--infrared, while $\phi \sim O(1)$ is a factor whose deviation from unity
can account for variations of the IMF from the standard Salpeter
(1955) $\psi(M) \propto M^{-2.35}$ form over the range 0.1 to 100 $M_{\odot}$
: for example, it should be 3.3 if a Miller--Scalo IMF is preferred, and
1/3.1 if the starburst forms only stars with $M > 1.6 M_{\odot}$. Cram et al. (1998)
presented a similar relationship (based on the ideas of Condon 1992), which reads
\begin{equation}
\frac {SFR (M \geq 5 M_{\odot})} {M_{\odot} \; \; {\rm yr}^{-1}} = \frac { L_{\nu}({\rm 60 \mu m})}{5.1 \times 10^{23} {\rm W Hz}^{-1}},
\label{eq:sfr5}
\end{equation}
assuming the same $\psi(M) \propto M^{-2.5}$ as in equation (\ref{eq:sfr1}).

Similar expressions can be derived in terms of the integrated far--infrared/submillimetre
emission. For example, the models of Devriendt et al. (1999) yield (for the same IMF as in
equations~\ref{eq:sfr2} and ~\ref{eq:sfr3}) the relationship
\begin{equation}
\frac {SFR} {M_{\odot} \; \; {\rm yr}^{-1}}  =  \left[ \frac {L_{\rm IR} (3-1000\mu {\rm m})}
{3.0 \times 10^{36} {\rm W}} \right]^{1.05}
\label{eq:sfr6}
\end{equation}
(cf. Eqns.~\ref{eq:sfr2} and~\ref{eq:sfr3} where the power law index deviates significantly from
unity).

As reviewed by Condon (1992) there is a well--known correlation between the far--infrared and 
decimetric radio luminosities of actively star--forming galaxies, which is thought to arise from
the fact  that the bulk of the radio luminosity is produced by synchrotron emission from 
relativistic electrons spiraling in the remnants of supernovae originating in the same population
of massive stars that produce the starburst component  to the far--infrared luminosity. Since the
cirrus component to the far--infrared luminosity is not expected to have associated radio emission,
it has been argued that the decimetric radio luminosity of a starburst galaxy may provide the cleanest
handle on its star formation rate, and Cram et al. (1998) present the following form for that 
relationship (assuming the same IMF as before):
\begin{equation}
\frac {SFR (M \geq 5 M_{\odot})} {M_{\odot} \; \; {\rm yr}^{-1}} = \frac {L_{\nu}({\rm 1.4 GHz})}{4.0 \times 10^{21}
{\rm W Hz}^{-1}},
\label{eq:sfr6} 
\end{equation}
where, as before, Cram et al. consider only the formation of stars with $5 \leq M \leq 100 M_{\odot}$ 
in an $M^{-2.5}$ IMF.
Additional SFR estimators exist, for example using the luminosity in the H$\alpha$ line (e.g. Kennicutt
1998), but these, too, are affected by dust, as shown by Rigopoulou et al. (2000) on the basis of
near--infrared VLT--ISAAC spectroscopy of the H$\alpha$ line in a subsample of the ISO--HDF--S objects
considered here.

\subsection{Comparing star formation estimates}
\label{sf_compare}

It is natural to enquire how these SFR estimators compare, and this question was addressed empirically
by Cram et al. (1998), using a somewhat heterogeneous compilation of data (U band magnitudes, 
H$\alpha$, 60$\mu$m and 1.4 GHz radio fluxes) from a variety of sources. They found that the star
formation rates deduced from the integrated far--infrared and decimetric radio luminosities are well correlated
over more than four orders of magnitude, but that the significant deviations from linearity and
greater scatter are seen for the relationships between the SFRs deduced from the 1.4 GHz power and
those from H$\alpha$ and U band luminosities. Some observational effects (e.g. slit losses in 
H$\alpha$ spectroscopy) may be contributing to these trends, but they are consistent with the
qualitative expectations of the picture outlined above in which massive stars are formed in dusty
environments, and reinforce the belief that this is the dominant mode of star formation in actively
star--forming galaxies.

Here we perform a complementary study, which attempts to circumvent the observational problems that
inevitably affect the analysis of any heterogeneous data sample drawn from many sources in the 
literature, by asking how these different star formation estimators fare when applied to model
SEDs for a range of types of star--forming galaxy. To do this we use the {\em GRASIL\/} models of 
Silva et al. (1998), which provide good fits to the UV--radio SEDs of six nearby galaxies, namely
 three starburst galaxies
(Arp220, M82 and NGC6090) with differing levels of activity, and three local spirals (M100, M51 and
NGC6946): similar models of starbursts are presented by Efstathious, Rowan--Robinson \& Siebenmorgen
(2000), but they do not consider normal spirals, which is why we use the {\em GRASIL\/} models
here. We refer the reader to the paper by Silva et al. (1998) for a discussion of the UV--mm
SEDs, and to Silva (1999) for details of their extension into the radio, through consideration of
separate thermal and non--thermal components to the radio emission, and note that Granato et al. 
(2000)
have also compared star formation estimators through their application to {\em GRASIL\/} SEDs. 

Silva et al. (1998) quote the
SFR value (averaged over the previous $5 \times 10^7$ yr) corresponding to each SED model, for an
assumed Salpeter (1955) IMF running from 0.1 to 100 $M_{\odot}$. Hence, to compare the SFR estimators, by
seeing how well each recovers those values, we 
must scale Eqns.(2) - (8) as required for them all to give the rate 
at which stellar mass would form given that reference IMF; this is analogous to the choice of 
the value of $\phi$ in the formalism of Rowan--Robinson et al. (1997). The derivation of the correct
scaling factor in each case requires consideration of the model through which the particular
observed luminosity is related to the rate of formation of massive stars: in  Eqns. (2), (3) \& (4) the 
value of a particular monochromatic UV luminosity is used as a direct tracer of these stars; underlying
Eqns. (5), (6), \& (7) is an assumption that some large ($\simeq 1$) fraction of their bolometric luminosity 
emerges [on a timescale of $\sim O(10^6$ yr)] in the far infrared (FIR), due to the reprocessing 
by dust; and the method of Eqn. (8) relates the number of high--mass stars that become Type Ib and Type II 
supernovae to the synchrotron emission produced by their remnants. 

In Appendix A, we show how each of these
physical models yields a scaling to be applied to Eqns. (2) - (8) to convert them to predictions for $\dot{M}$,
the rate of formation of stellar mass, given our canonical IMF, which is a Salpeter (1955) IMF running from 0.1 
to 100 $M_{\odot}$, in our canonical cosmology, which is an Einstein -- de Sitter universe with a Hubble
constant of 50 km s$^{-1}$ Mpc$^{-1}$. We find the following set of corrected versions of the estimators
of eqns. (2) to (8):

\begin{equation}
\frac {\dot{M}}{M_{\odot} {\rm yr}^{-1}}  = \left\{ \begin{array}{lllllll}

 L_{\nu}(U) / 2.6 \times 10^{21} {\rm W Hz}^{-1}  \\
\left[ \lambda L_{\lambda}(280{\rm nm}) /7.0 \times 10^{34} {\rm W} \right]^{1.62} \\ 
\left[\lambda L_{\lambda}(160{\rm nm}) / 8.4 \times 10^{34} {\rm W} \right]^{1.72}  \\ 
\lambda L_{\lambda}(60 \mu {\rm m}) / 1.5 \times 10^{36} {\rm W}\\
L_{\nu}(60 \mu {\rm m}) / 1.1 \times 10^{23} {\rm W Hz}^{-1} \\
\left[ L_{\rm IR}(3-1000 \mu {\rm m}) / 2.3 \times 10^{36} {\rm W} \right]^{1.05} \\
L_{\nu}(1.4 {\rm GHz}) / 6.9 \times 10^{20} {\rm W Hz}^{-1}. \\
						    \end{array} \right.
\label{eq:scaled_sfr}
\end{equation}	

Having made these various scalings we are ready to compare the SFR values obtained by applying the
estimators of eqn.~\ref{eq:scaled_sfr} to the six {\em GRASIL\/} SEDs with the values
given for them by Silva et al. (1998), as shown in Figure~\ref{fig:sfr_methods}: note that 
the results from the two 60$\mu$m--based estimators are plotted together.
The first thing to note from this figure is that the far infrared and radio estimators do far better at
recovering the {\em GRASIL\/} SFRs than do the three UV--based ones. The integrated far infrared
estimator of Devriendt et al. (1999) is the closest to the {\em GRASIL\/} SFR value for all models,
while the 1.4 GHz is offset by a fairly constant factor of $\sim3$, indicating that, while there may
be a problem with its absolute normalisation, (possibly caused by the way that the {\em GRASIL\/}
SEDs are extended into the radio), this method works well for all the SED types considered,
yielding accurate relative SFR values for them. The 60$\mu$m--based estimators fare quite well, too,
although the ratio of their SFR values to those from the {\em GRASIL\/} code are higher for starbursts
(filled symbols) than for normal galaxies (empty symbols), suggesting that a far infrared colour term 
should be included in the relationship between SFR and far infrared luminosity.

The three UV--based estimators display a much more complex
behaviour across the range of galaxy types. 
The U band luminosity always gives a lower SFR estimate than the four integrated/far--infrared and radio
estimators, and the discrepancy is appreciably larger for the starbursts than for the normal
spirals, as expected if the UV light is generated in dustier environments in starbursts than in normal
spirals. The two non--linear UV--SFR relations deduced by  Devriendt et al. (1999)
appear to over--estimate the star formation rates of the spirals, and one of the starbursts
(NGC6090), but do reasonably well for M82 and, especially, Arp220. This could be because the 
Devriendt et al. (1999) models, although spanning the full range of galaxy types from inactive spirals to
ULIRGs, are more directed at the understanding of starbursts and/or could simply reflect that the
best--fit correlations yielding Eqns.(3) and (4) do not express anything
physical in the models.

It is clear from these results that caution must be exercised in comparing SFRs in the literature,
which might have been made using different estimators and with differing assumptions as to the 
stellar IMF and the exact specification of the astrophysical model underlying their application.
Even when scaled to a common reference IMF, the SFR estimators based on far infrared or radio
luminosities are only consistent to within a factor of two, while the UV--based ones are seen
to be far less secure. The consistency between the ratios of SFR
values obtained by the four radio-- and integrated/FIR--based estimators across the six SEDs is, of course, simply a
manifestation of the universality of the radio--FIR correlation (reviewed by Condon 1992), but our
results do afford some confidence that they are measuring quantities 
correlated with the ``true'' star formation rate, at least to the extent that is well reproduced
by the {\em GRASIL\/} models, which cannot be said for the UV--based estimators. On the basis of these
results, the integrated IR luminosity appears to be the best SFR estimator out of the set we have
investigated, so it is that which we shall use in what follows, when we deduce SFR values for our
ISO sources.

\section{Spectral energy distributions and star formation rates of {\em ISO}--HDF--S sources}
\label{sect:sed_sfr}

\begin{figure*}
\caption{Spectral energy distributions for the 24 objects from 
Table~\ref{tab:id_props} identified as galaxies. The squares mark the
UV/optical/near--IR and radio
photometric data for the objects, from Table~\ref{tab:id_props} while the solid, dotted,
dashed, dash--dot, dash--dot--dot--dot and long--dashed lines show,
respectively, the GRASIL model fits to the SEDs of Arp220, M100, M51,
M82, MGC6090 and NGC6946, redshifted as appropriate for each galaxy.
The {\em ISO} data from Table 8 of Paper I are plotted as error bars or
upper limits, as appropriate.
{\bf High resolution version of this figure available from astro.ic.ac.uk/hdfs.}}
\label{fig:seds}
\end{figure*}

\begin{figure*}
\contcaption{Spectral energy distributions for the 24 objects from 
Table~\ref{tab:id_props} identified as galaxies. The squares mark the
UV/optical/near--IR and radio
photometric data for the objects, from Table~\ref{tab:id_props} while the solid, dotted,
dashed, dash--dot, dash--dot--dot--dot and long--dashed lines show,
respectively, the GRASIL model fits to the SEDs of Arp220, M100, M51,
M82, MGC6090 and NGC6946, redshifted as appropriate for each galaxy.
The {\em ISO} data from Table 8 of Paper I are plotted as error bars or
upper limits, as appropriate.
{\bf High resolution version of this figure available from astro.ic.ac.uk/hdfs.}}
\end{figure*}

\begin{figure*}
\contcaption{Spectral energy distributions for the 24 objects from 
Table~\ref{tab:id_props} identified as galaxies. The squares mark the
UV/optical/near--IR and radio
photometric data for the objects, from Table~\ref{tab:id_props} while the solid, dotted,
dashed, dash--dot, dash--dot--dot--dot and long--dashed lines show,
respectively, the GRASIL model fits to the SEDs of Arp220, M100, M51,
M82, MGC6090 and NGC6946, redshifted as appropriate for each galaxy.
The {\em ISO} data from Table 8 of Paper I are plotted as error bars or
upper limits, as appropriate.
{\bf High resolution version of this figure available from astro.ic.ac.uk/hdfs.}}
\end{figure*}

\begin{figure*}
\contcaption{Spectral energy distributions for the 24 objects from 
Table~\ref{tab:id_props} identified as galaxies. The squares mark the
UV/optical/near--IR and radio
photometric data for the objects, from Table~\ref{tab:id_props} while the solid, dotted,
dashed, dash--dot, dash--dot--dot--dot and long--dashed lines show,
respectively, the GRASIL model fits to the SEDs of Arp220, M100, M51,
M82, MGC6090 and NGC6946, redshifted as appropriate for each galaxy.
The {\em ISO} data from Table 8 of Paper I are plotted as error bars or
upper limits, as appropriate.
{\bf High resolution version of this figure available from astro.ic.ac.uk/hdfs.}}
\end{figure*}

\begin{figure*}
\contcaption{Spectral energy distributions for the 24 objects from 
Table~\ref{tab:id_props} identified as galaxies. The squares mark the
UV/optical/near--IR and radio
photometric data for the objects, from Table~\ref{tab:id_props} while the solid, dotted,
dashed, dash--dot, dash--dot--dot--dot and long--dashed lines show,
respectively, the GRASIL model fits to the SEDs of Arp220, M100, M51,
M82, MGC6090 and NGC6946, redshifted as appropriate for each galaxy.
The {\em ISO} data from Table 8 of Paper I are plotted as error bars or
upper limits, as appropriate.
{\bf High resolution version of this figure available from astro.ic.ac.uk/hdfs.}}
\end{figure*}

\begin{figure*}
\contcaption{Spectral energy distributions for the 24 objects from 
Table~\ref{tab:id_props} identified as galaxies. The squares mark the
UV/optical/near--IR and radio
photometric data for the objects, from Table~\ref{tab:id_props} while the solid, dotted,
dashed, dash--dot, dash--dot--dot--dot and long--dashed lines show,
respectively, the GRASIL model fits to the SEDs of Arp220, M100, M51,
M82, MGC6090 and NGC6946, redshifted as appropriate for each galaxy.
The {\em ISO} data from Table 8 of Paper I are plotted as error bars or
upper limits, as appropriate.
{\bf High resolution version of this figure available from astro.ic.ac.uk/hdfs.}}
\end{figure*}

In this Section we estimate star formation rates for our {\em ISO\/} sources, through the
application of the integrated IR luminosity estimator judged to be the best in 
Section~\ref{sf_compare} to {\em GRASIL\/} model SEDs fitted to the photometric data for
each of the 24 galaxies given in Table~\ref{tab:id_props}, plus radio data (mostly upper limits) kindly 
provided in advance of publication by the ATNF HDF--S survey team (A. Hopkins, {\em priv. comm.}).
We do not expect  our {\em ISO\/} sources to match exactly one of the six {\em GRASIL\/} models,
but, to the extent that the models span the range of SEDs of star--forming galaxies
likely to feature in our mid--infrared survey, this method gives us a handle on the uncertainty
in the deduced SFR value of each {\em ISO\/} source resulting from uncertainty in its true SED:
note that the SEDs of the three normal spirals are very similar, and it is only when an appreciable
starburst component kicks in that the shape of the SED begins to change significantly.
To facilitate the choice of the {\em GRASIL\/} model most appropriate to each galaxy, we
plot in Figure~\ref{fig:seds} the photometric data set for each source, together with the six model
SEDs, normalized in each case to match the K band magnitude of the galaxy associated with the
ISO--HDF--S source, or the I band magnitude in the one case (J223240-603141) of a galaxy lying outside
the region of the EIS K band survey. From this figure we conclude the following about the 
star formation rates of the 24 ISO--HDF--S galaxies:

\begin{enumerate}

\item {\bf ISOHDFS~J223240-603141:} The observed I band to 15$\mu$m colour suggests that this is a starburst
galaxy, rather than a normal spiral dominated by cirrus emission: the lack of a detection at 7$\mu$m
is just consistent with that, and J223240-603141 does lie close to a very bright 7$\mu$m source, so it
is possible its 7$\mu$m flux estimate has been corrupted by the subtraction of the negative lobes
of the bright source. The lack of detections at 1.4, 2.5 and 4.9 GHz argue against the
starburst being as extreme as that in Arp220, and the optical colours favour an SED like that of
NGC6090 over that of M82. Using the 1.4GHz, 60$\mu$m (Cram et al.), 60$\mu$m (Rowan--Robinson et al.,
1997, with $\epsilon=\phi=1$), 3-1000$\mu$m prescriptions we obtain SFR estimates of 43, 34, 12 and 20 
$M_{\odot} \; {\rm yr}^{-1}$, respectively, from adopting this SED. On the basis of 
Section~\ref{sf_compare}, we adopt the last of these, so our best estimate of the true SFR for 
this galaxy is 20$M_{\odot} \; {\rm yr}^{-1}$. 

\item {\bf ISOHDFS~J223243-603242:} This is a slightly ambiguous case. The optical/near--infrared colours of this
galaxy fit the M82 model SED very well, as does the 7$\mu$m flux, once the model has been normalized to
the observed K band flux of J223243-603242; however the 15$\mu$m flux seems a little low and the
lack of a 1.4 GHz detection is marginally inconsistent with the M82 model. Using the M82 model
SED, the same four SFR estimators give 145, 121, 43, 60 $M_{\odot} \; {\rm yr}^{-1}$, while, if
we had adopted an M100 SED instead, we would have obtained values of 22, 9, 3 and 7$M_{\odot} \; 
{\rm yr}^{-1}$, respectively. The M82 SED is a better fit overall, so we adopt a best guess 
SFR value for this galaxy of  60 $M_{\odot} \; {\rm yr}^{-1}$, noting that, while the precise SFR value
is uncertain, it is clear this galaxy is forming stars at a rate of several tens of $M_{\odot} \; 
{\rm yr}^{-1}$.

\item {\bf ISOHDFS~J223243-603351:} This is one of the four ISO--HDF--S sources detected in the radio and the
strength of these detections (at 4.9 and 8.6 GHz), together with the optical/near--infrared SED of the
galaxy, which rises into the UV, suggests that this may be an AGN, and this is confirmed by the
presence of a broad line in its optical spectrum: we therefore, do not estimate the SFR in this source.

\item {\bf ISOHDFS~J223243-603441:} This is a second source, like J223243-603242, which has an 
optical/near--infrared SED well matching the starburst models, but {\em ISO} fluxes which do not
unambiguously support that interpretation, as the 15$\mu$m flux is lower than would be expected for
a starburst on the basis of its SED up to 6.7$\mu$m. From Fig.~\ref{fig:id_post} we see that this
source is close to a much brighter source at 15$\mu$m, so it is possible that its flux in that
band has been under-estimated, due to the difficulty of subtracting the negative lobe from the
brighter source. If we assume that, and adopt an NGC6090 SED, then we obtain SFR values of 216, 169, 
61, 108  $M_{\odot} \; {\rm yr}^{-1}$ from the usual four estimators, while, if we take the 
6.7$\mu$m--15$\mu$m colour at face value, favouring an M51 SED, we obtain, instead, values of 
49, 28, 10 and 19  $M_{\odot} \; {\rm yr}^{-1}$. As a compromise, we adopt a best guess of 50 
 $M_{\odot} \; {\rm yr}^{-1}$, but note that is is uncertain by a factor of two, at least.

\item {\bf ISOHDFS~J223244-603110:} The optical  to near--infrared SED of this galaxy is a little unusual, 
suggesting
a mis--match in the apertures used to measure the optical and near--infrared magnitudes. Fixing
the model SEDs to match the K band magnitude we obtain a good fit to the M82 and NGC6090 models, yielding
an SFR estimate of 4 $M_{\odot} \; {\rm yr}^{-1}$ from applying the integrated IR luminosity
estimator to the fitted M82 SED mode, so this is not a very strong starburst.

\item {\bf ISOHDFS~J223244-603455:} The proximity of a brighter 7$\mu$m source might explain the lack of a
detection in that band, due to the lobe--subtraction problem, but it seems more likely, on all evidence,
that this is a normal spiral and not a starburst. Using the NGC6946 model SED the integrated $L_{\rm 
IR}$ estimator yields 4$M_{\odot} \; {\rm yr}^{-1}$, a modest star formation rate, consistent 
with this interpretation.

\item {\bf ISOHDFS~J223245-603226:} Again, the lack of a 7$\mu$m detection slightly confuses an otherwise fairly
confident identification of this with an M82- or NGC6090--type starburst, and, once more, this source
is close to a strong 7$\mu$m source, so this may be due to the lobe--subtraction problem. If we adopt
the NGC6090 SED model, which gives a slightly better fit to the optical/near--infrared SED, 
we obtain a best guess SFR of 36$M_{\odot} \; {\rm yr}^{-1}$ for this galaxy.

\item {\bf ISOHDFS~J223245-603418:} This is one of the most unambiguous starburst detections. Its UV--15$\mu$m
colours are well fitted by the M82 SED (and almost as well by the Arp220 model), but the solid radio 
detections at 1.4, 2.5 and 4.9 GHz suggest a starburst stronger than M82. Using the M82 and Arp220 SEDs 
we obtain SFRs of 40 and 300$M_{\odot} \; {\rm yr}^{-1}$, using the integrated $L_{\rm IR}$ estimator, 
confirming that this is a powerful starburst, although leaving some doubt
as to its true star formation rate: we adopt a best guess of 100 $M_{\odot} \; {\rm yr}^{-1}$, but
note this is uncertain by a factor of two, at least.

\item {\bf ISOHDFS~J223247-603335:} Another case where an unambiguous discrimination between spiral and starburst
is difficult: the {\em ISO} fluxes are a little lower than expected for the starburst fit to the 
optical/near--infrared SED. Taking the NGC6090 and M100 SEDs, which bracket the data points, we obtain
star formation rates of 130 and 16 $M_{\odot} \; {\rm yr}^{-1}$, from application of the integrated
 $L_{\rm IR}$ estimator, so there is quite some uncertainty in the SFR for the source: we adopt a value
of 50$M_{\odot} \; {\rm yr}^{-1}$, again with a factor of two uncertainty, at least, although it is
clear that this galaxy is definitely forming stars actively.

\item {\bf ISOHDFS~J223251-603335:} The SED of this source is quite well fit by the NGC6090 model, with which
the integrated  $L_{\rm IR}$ estimator yields an SFR estimates of 10 $M_{\odot} \; {\rm yr}^{-1}$, so
it is clear this is a modest starburst.

\item {\bf ISOHDFS~J223252-603327:} This radio source, detected solidly at 1.4 GHz, is at the centre of extended
emission in both {\em ISO} bands, which may explain why the mid--infrared fluxes slightly exceed the predictions
of the starburst models. The radio detection is in excellent agreement with the Arp220 model, and if
we adopt that we obtain 500 $M_{\odot} \; {\rm yr}^{-1}$, using the integrated $L_{\rm IR}$ estimator.
It was noted in Subsection~\ref{sect:id_notes} that this source becomes more point-like at longer 
wavelengths, as one passes through the optical into the near--infrared, so it could contain an obscured 
AGN, and, hence, the star formation rate estimated here may be an over--estimate.

\item {\bf ISOHDFS~J223254-603115:} As noted in Section~\ref{sect:id_notes}, there appears to be a mismatch in 
the apertures used for the optical and near--infrared photometry of this galaxy. If we normalize the  
{\em GRASIL\/}  models to the K band magnitude of this galaxy, then we see that its SED is in good 
agreement with the NGC6090 and M82 models, with the Arp220 model marginally
disfavoured by the lack of a radio detection. Applying the integrated $L_{\rm IR}$ estimator to the
NGC6090 model we obtain an SFR of 11$M_{\odot} \; {\rm yr}^{-1}$, which we adopt as our best guess; 
if, instead, the M82 model were used, that
figure would be negligibly higher, at 12  $M_{\odot} \; {\rm yr}^{-1}$.

\item {\bf ISOHDFS~J223254-603129:} From Figure~\ref{fig:seds} we see that the
{\em ISO} fluxes of this source are in 
good agreement with the three starburst model SEDs, when they are normalized using the galaxy's K band
magnitude. If we apply the integrated  $L_{\rm IR}$ estimator to the NGC6090 model, which gives the best
overall fit, we obtain a best guess SFR estimate of 1.2 $M_{\odot} \; {\rm yr}^{-1}$.

\item {\bf ISOHDFS~J223257-603305:} The 15$\mu$m to near--IR
colour of this galaxy implies that it is a starburst, and the lack of radio detections excludes a burst
as extreme as Arp220. Taking the NGC6090 model, we obtain an SFR estimate of 43
$M_{\odot} \; {\rm yr}^{-1}$, using the integrated $L_{\rm IR}$ estimator, and we adopt this as our best guess.

\item {\bf ISOHDFS~J223302-603323:}  Figure~\ref{fig:seds} shows that this source is securely identified as a 
cirrus galaxy, rather than starburst, and, adopting the M100 model, we obtain a best guess SFR of 
10 $M_{\odot} \; {\rm yr}^{-1}$, using the integrated $L_{\rm IR}$ estimator.

\item {\bf ISOHDFS~J223303-603336:} The 7$\mu$m to near--IR colour of this source suggests that it is a starburst,
rather than a cirrus galaxy, 
but the lack of a 15$\mu$m detection is puzzling, in that case, although once again
it may be due to uncertainty in the subtraction of the negative lobes from a nearby source that is
bright in both {\em ISO\/} bands. The
Arp220 SED is ruled out by the lack of radio detections, and, if we adopt the NGC6090 model, we obtain 
an  SFR of 16$M_{\odot} \; {\rm yr}^{-1}$, using the integrated $L_{\rm IR}$ estimator, while, if the M100
model is preferred, this falls to 2$M_{\odot} \; {\rm yr}^{-1}$. We shall adopt a best guess of 
5 $M_{\odot} \; {\rm yr}^{-1}$, but note that this is one of our most uncertain SFR estimates. 

\item {\bf ISOHDFS~J223306-603349:} From  Figure~\ref{fig:seds} we see that this galaxy was detected in the radio
at 1.4 and 2.5 GHz at almost exactly the level expected for the {\em GRASIL\/} NGC6090 SED normalized 
to the galaxy's observed K band magnitude, but that the two {\em ISO} fluxes look a little low to match 
that interpretation, especially the 15$\mu$m flux. The radio detections are a strong indication that 
this is a starburst, rather than a cirrus galaxy, however, so we do adopt  the NGC6090 SED, which 
yields a best guess SFR of  60$M_{\odot} \; {\rm yr}^{-1}$, from the integrated $L_{\rm IR}$ estimator,
although we record that, if we had adopted a NGC6946 SED, this would have fallen to 10 
 $M_{\odot} \; {\rm yr}^{-1}$.

\item {\bf ISOHDFS~J223306-603436:} The mid--IR colour of this source suggests that it is a starburst, rather 
than a cirrus galaxy, and its optical/near--IR colours fit an (appropriately normalized) M82 model all 
the way from the B to K band. Adopting that SED yields a best guess SFR of 74$\pm$15 $M_{\odot} \; 
{\rm yr}^{-1}$. 

\item {\bf ISOHDFS~J223306-603450:} As illustrated in Figure~\ref{fig:seds}, the 15$\mu$m/near--IR colour of 
this source is that of a starburst, not a cirrus galaxy, but the extant data are not able to distinguish
between an Arp220--type galaxy and a more modest burst, like M82, with the lack of detections at 7$\mu$m 
and in the radio consistent with both possibilities. The Arp220 SED would yield an SFR of 130 
 $M_{\odot} \; {\rm yr}^{-1}$, while the M82 model gives  17$M_{\odot} \; {\rm yr}^{-1}$, so we shall
adopt an admittedly uncertain  compromise figure of 50$M_{\odot} \; {\rm yr}^{-1}$ as our best guess SFR.

\item {\bf ISOHDFS~J223307-603248:} We cannot unambiguously distinguish whether this source is a starburst or 
cirrus galaxy. The 
7/15$\mu$m colours favours the former, as does the shape of its UV/optical/near--IR SED, and, if we 
adopt the NGC6090 SED we obtain an SFR estimate of 65$M_{\odot} \; {\rm yr}^{-1}$, while a 
lower value of 10 $M_{\odot} \; {\rm yr}^{-1}$ would result from using an M100 model instead: we shall
adopt the former value as our best guess, but note that we cannot confidently exclude an SFR value
a factor of four lower.

\item {\bf ISOHDFS~J223308-603314:} This source is solidly identified as a starburst, as its SED fits the 
{\em GRASIL\/} NGC6090 model from the U band all the way to 15$\mu$m. With that SED we estimate its 
SFR to be 46$M_{\odot} \; {\rm yr}^{-1}$, using the integrated $L_{\rm IR}$ estimator.

\item {\bf ISOHDFS~J223312-603350:} As noted in Section~\ref{sect:id_notes}, the redshift of this galaxy is
highly uncertain, and, hence, so is its star formation rate. The shape of its UV/optical/near--IR SED
suggests that it is a starburst, as, more strongly, does its mid/near--IR colour, but it is difficult
to judge the strength of its burst from the extant data. We adopt an NGC6090 SED to obtain a best
guess SFR estimate of 15 $M_{\odot} \; {\rm yr}^{-1}$, although we note that choosing an Arp220
model instead would have yielded 130 $M_{\odot} \; {\rm yr}^{-1}$.

\item {\bf ISOHDFS~J223312-603416:} This is another source with a very uncertain redshift, based only on JHK
photometry, and, furthermore, its 7$\mu$m flux is highly uncertain, due to its being located in a 
region of extended 7$\mu$m emission, as shown in Figure~\ref{fig:id_post}.
The 15$\mu$m/near--IR colour of this source suggests that it is a starburst, and the
Arp220 and NGC6090 SEDs give SFR estimates of 370 and 46 $M_{\odot} \; {\rm yr}^{-1}$,
respectively. We adopt a value of 100 $M_{\odot} \; {\rm yr}^{-1}$ as our best guess, but noting that,
in addition to the uncertainty due to the poorly constrained SED, there is also uncertainty over
the redshift of this galaxy.

\item {\bf ISOHDFS~J223314-603203:} Another source whose mid/near--IR colour indicates it to be a starburst,
rather than a cirrus galaxy, but for which we are unable to assess the strength of the burst given the 
current data. The Arp220 and NGC6090 SED models give SFR estimates of 9 and 1 $M_{\odot} \; 
{\rm yr}^{-1}$, respectively, indicating either way that this is a modest starburst: we shall
adopt a value of 3 $M_{\odot} \; {\rm yr}^{-1}$ as our best guess, noting that Section 3 showed this to be
one of our most uncertain associations.

\end{enumerate}

\begin{figure}
\hspace{-0.7cm}
\epsfig{file=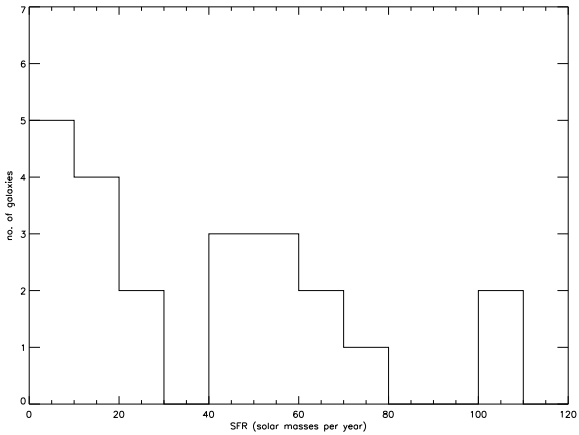,width=6cm,angle=90}
\caption{The distribution of star formation rates for the 22 {\em ISO}
HDF--S sources we believe to be star
forming galaxies, omitting the galaxy with the highest SFR (ISOHDFS~J223252-603327 at 500$M_{\odot} \; 
{\rm yr}^{-1}$), which may be harbouring an obscured AGN.}
\label{fig:sfrhist}
\end{figure}

\begin{figure}
\hspace{-0.7cm}
\epsfig{file=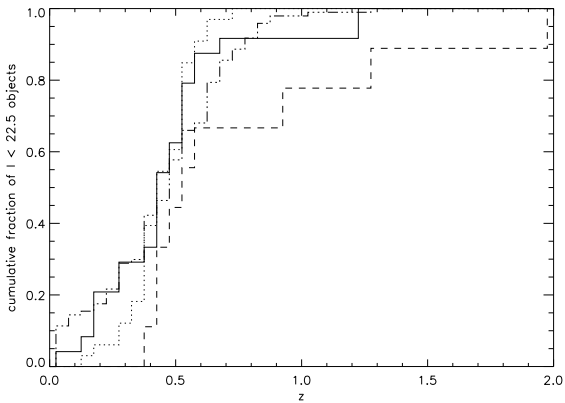,width=8cm}
\caption{A comparison of the cumulative redshift distributions of {\em
ISO}--HDF--S galaxy sample (solid line)
with those from three HDF--S photometric redshift catalogues: the dot-dot-dot-dashed line is that of MRR,
the dashed line that of Gwyn, and the dotted line that of the SUNY group.} 
\label{fig:zdist}
\end{figure}

\section{The contribution of {\em ISO\/}--selected sources to the star formation history of the HDF--S}
\label{sect:madau}

In Figure~\ref{fig:sfrhist} we plot the distribution of SFR values for
22 {\em ISO}--HDF--S sources we believe to be star--forming
galaxies, omitting the galaxy with the highest SFR (ISOHDFS~J223252-603327 at 500$M_{\odot} \; 
{\rm yr}^{-1}$), which may be harbouring an obscured AGN. As can be seen from this, the ISO--HDF--S 
galaxies are typically active starbursts (median SFR = 43 $M_{\odot} \; {\rm yr}^{-1}$), although 
they do cover a range of two orders of magnitude in star formation rate. These values are
computed under the assumption of an Einstein -- de Sitter Universe with a Hubble constant of 
$H_0=50$~km~$s^{-1}$~Mpc$^{-1}$. Using the  slightly non--linear integratd $L_{\rm IR}$ estimator of Devriendt 
et al. (1999) we have the SFR varying with luminosity distance, $d_{\rm L}$, as $d_{\rm L}^{2.1}$,
and, hence, with Hubble constant as $H_0^{0.48}$. If, instead, we had assumed a cosmology with
$\Omega_{\rm M}=0.7$ and $\Omega_{\Lambda}=0.3$, we would have deduced, for the same value of
$H_0$ SFR values that were factors of (11, 20, 47 \& 78) per cent higher for galaxies at redshifts
of (0.1, 0.2, 0.5 \& 1.0) respectively.

For five of our galaxies, SFR estimates have been made by Rigopoulou et al. (2000), on the basis of
H$\alpha$ luminosity and also using a far--infrared (FIR) estimator (Franceschini et al., in preparation)
which makes use of the 15$\mu$m flux and assumed a far--infrared/mid--infrared luminosity ratio of
$\sim10$ as appropriate for a galaxy with a spectral energy distribution like that of M82. Rigopoulou
et al. (2000) assumed a cosmology with $\Omega_{\rm M}=0.7$ and $\Omega_{\Lambda}=0.3$ and a 
Salpeter (1955) IMF over the mass range [1,100]$M_{\odot}$, so, before we can compare their results
with ours, we must scale their SFR values to account for these effects, using methods discussed 
in Appendix A.
Under our canonical assumptions, their FIR--based estimator gives SFR values of (113, 89, 69, 220, 65)
$M_{\odot} {\rm yr}^{-1}$ for galaxies (J223245-603418, J223245-603226, J223247-603335, J223252-603327,
J223257-603305), respectively, for which our adopted values are (100, 36, 50, 500, 43)$M_{\odot} 
{\rm yr}^{-1}$ , which is a reasonable level of agreement, given the uncertainties in the true SEDs
of these sources. With the appropriate scalings, the H$\alpha$--based estimator of Rigopoulou et al.
(2000) gives SFR raw values of (2.1, 6.0, 5.9, 74, 12) $M_{\odot} {\rm yr}^{-1}$ for these five galaxies,
which are factors of 5--50 smaller than those estimated from the FIR. Rigopoulou et al. (2000) argue
that, on the basis of the V-K colours of these galaxies, one would deduce an extinction correction of
only $\sim 4$ to the H$\alpha$ SFR, suggesting that optical data alone are not sufficient to derive
a good SFR value, even when they do lead an extinction correction; this is in accordance with the work
of Silva et al. (1998) and Jimenez et al. (1999), mentioned above, who argue on theoretical grounds
that the emission from star--forming regions cannot be well represented by 
a simple model of a dust--screen in front of a dust--free starburst. 

\begin{figure*}
\hspace{-0.7cm}
\epsfig {file=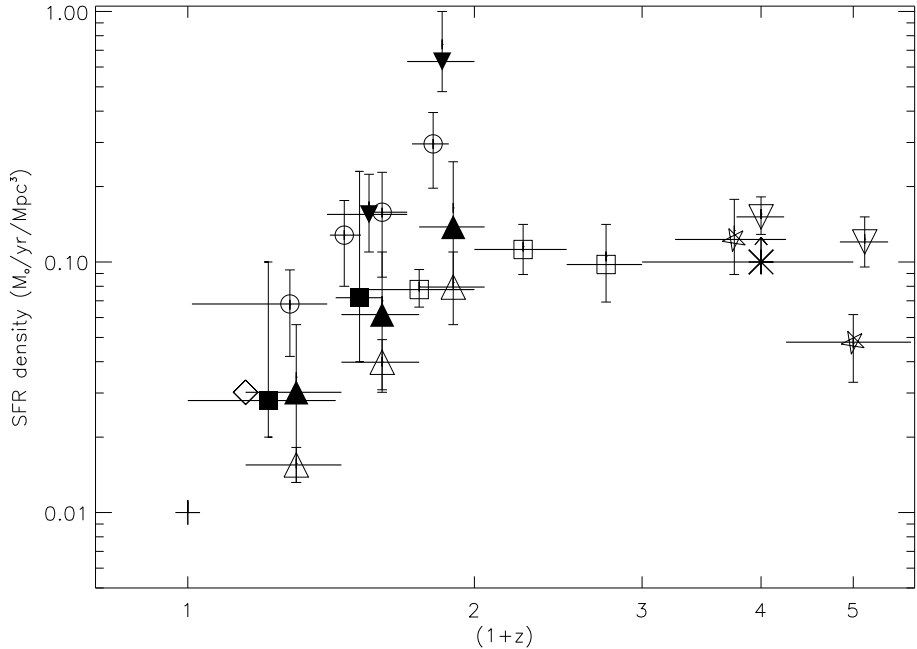,width=13cm,angle=0}
\caption{A compilation of constraints on the star formation history of the Universe, mainly taken
from Haarsma et al. (2000): in all cases, an Einstein -- de Sitter universe, with a Hubble constant
of 50 km s$^{-1}$ Mpc$^{-1}$ was assumed, as was star formation taking place with a Salpeter (1955)
IMF, over the mass range $[0.1,100] M_{\odot}$, and the dust extinction corrections of Steidel et al. 
(1999) were applied. The symbols are as follows: asterisk -- Hughes et al. (1998); filled upward--pointing
triangle -- Flores et al. (1999); empty square -- Connolly et al. (1997); plus sign -- 
Gallego et al. (1995); empty downward--pointing triangle -- Steidel et al. (1999); five--pointed star -- 
Madau et al. (1996); diamond --
Treyer et al. (1998); filled downward--pointing triangle -- Rowan-Robinson et al. (1997); empty 
upward--pointing triangle -- Lilly et 
al. (1996); empty circle -- Haarsma et al. (2000); filled square -- this work, with confidence
intervals computed according to the sampling variance treatment of Appendix B, assuming $\sigma^2=1$.}
\label{fig:madau}
\end{figure*}

\subsection{Computing the raw star formation rate density}
\label{sect:raw_madau}

We may use these SFR estimates to derive constraints on the star formation history of the Universe,
at least in the redshift interval in which our sources are found, i.e. $z \leq 0.6$ for the most
part. In fact, as Figure~\ref{fig:zdist} shows, the redshift distribution of the galaxies associated with
our {\em ISO\/} sources is consistent with that of similarly bright (i.e. $I < 22.5$) galaxies
in the HDF--S as a whole, at least as judged from photometric redshift catalogues produced
for the field: two--sided Kolmogorov--Smyrnov tests reveal that the cumulative redshift distribution 
of the {\em ISO\/} IDs has probabilities of 0.30, 0.43 and 0.75 of being drawn from the same
population as that yielding  the $n(z)$ distributions found in the photometric redshift
catalogues of MRR, Gwyn and the SUNY group, respectively.

We divide this region into two equal volume bins ($0 \leq z \leq 0.43$ and $0.43 \leq z \leq
0.6$), in which we compute the star formation rate density, via the available volume
technique.  The volume in which each galaxy could have been observed within any
given redshift slice was
estimated by considering its luminosity and best fit SED and hence determining
the effective area over which it could have been observed at given redshift
using the area as a function of 15$\mu$m flux limits (Figure 13 of Paper I).
The contribution of each galaxy to the global star formation rate density, $\dot{\rho}_*$,
was then calculated as the ratio between the star formation rate 
determined for that galaxy and its available volume, and the sum of these individual 
contributions was then taken over all galaxies in each bin. Note that there exists one galaxy
(ISOHDF~J223303-603336, which may be affected by the lobe subtraction problem) lying the range 
$z<0.6$ which does not have a detection at 15$\mu$m. Whether
this galaxy is excluded, or included, using an available volume computed with an estimate of 
what its 15$\mu$m flux ``should'' be, on the basis of the best fitting GRASIL SED, makes a
negligible difference to the raw value of  $\dot{\rho}_*$ obtained, so, for definiteness, we
neglect it, leaving us with a 15$\mu$m--selected sample.

This yielded values of $\dot{\rho}_* = 0.03 \pm 0.2 {\rm dex}$ ({\em i.e.\/} $\dot{\rho}_* = 0.03^{+0.02}_{-0.01}$)
and $\dot{\rho}_* = 0.07 \pm 0.1 {\rm dex}$ ({\em i.e.\/} $\dot{\rho}_* = 0.07^{+0.02}_{-0.01}$)
$ M_{\odot} {\rm yr}^{-1} {\rm Mpc}^{-1}$ for $0 \leq z \leq 0.43$ and 
$0.43 \leq z \leq 0.6$, respectively, where we have assumed an Einstein -- de Sitter Universe,  star formation
according to a Salpeter (1955) IMF over the mass range [0.1,100] $M_{\odot}$ and taken our SFR estimates
for individual galaxies to be uncertain by a factor of two. In Figure~\ref{fig:madau}
we show a comparison of these results and a compilation (principally
that of Haarsma et al. 2000) of similar constraints
derived recently from a variety of methods in a number of wavebands, all using the same 
assumed cosmology and IMF. One notable feature of this is 
that our raw results are in excellent agreement with another recent study using {\em ISO\/} data, that
of Flores et al. (1999) based on  their study of the CFRS 1415+52 field. 

\subsection{Sampling variance in our $\dot{\rho}_*$ results}\label{sect:sampling}

The relatively small volume of our survey region means that our star formation
rate density values are subject to a relatively large sampling variance.
The volume of a cone $0<z<0.43$ with solid angle 20 sq. arcmin (equal to our survey area) is 
$\sim$520 $h^{-3} {\rm Mpc}^3$, while a frustum with the same solid angle and
$0.43<z<0.6$ is $\sim$570 $h^{-3} {\rm Mpc}^3$.  We can thus consider each
survey section to have a volume of $\sim 550h^{-3}$Mpc$^3$; cubic cells of the
same volume would have sides of 8.2 $h^{-1}$ Mpc, while spheres would have radius of
5.0 $h^{-1}$ Mpc. These volumes are small enough that the variation in mean
density between surveys carried out in different regions of the Universe are
expected to be significant.  Using the power spectrum of Peacock \& Dodds (1994)
and the conversion from length scale to effective wavenumber
for cubic cells (Peacock 1991)  we estimate that the fractional
density fluctuations in cubic cells of this volume at redshift zero will be $\sigma^2=1.05$,
while that of spheres would have  $\sigma^2=1.15$. Note that the variances in local cells with the
same geometry as our survey volumes are likely to be larger than this, as their elongated
shape would lead to a window function that lets in more small--scale power than spherical
or cubic window functions do. Several more effects further complicate an assessment of the sampling
variance in our SFR density estimates. Firstly, there is evolution in the density field with
redshift across the two sampling volumes.  At $z=0.2$ and $z=0.5$, in the middle of our two redshift 
bins, the linear theory variances in a given volume would be factors of $(1+z)^2$, i.e. 1.4 and 2.3, 
respectively, below their values at redshift zero, while non--linear effects as $\sigma \rightarrow 1$
are likely to make the decrease more pronounced than these linear theory estimates suggest. Furthermore,
the shape of our survey volumes means that
they have more volume at higher redshift, where the density field is less evolved. A further 
complicating factor is the bias between the galaxy and mass distributions, which is likely to lead to
a higher sampling variance for the total SFRs in our two redshift bins than that for the mass they
contain, and an additional uncertainty in the redshift variation across the bins: Kauffmann et al.
(1999) show that the evolution of galaxy clustering strength over \mbox{$0 \leq z \leq 1$} is
a function of cosmology, at least in  their galaxy formation models.
Clearly, the quantitative assessment of these various factors is beyond the scope of the
current paper, so we shall base this study of the effects of sampling variance on   $\dot{\rho}_*$
estimates on the assumption that the variance in the mass contained in our two redshift bins
is $\sigma^2 \sim 1$ and that $\sigma^2 =0.5$ and $\sigma^2 = 2$ are likely to be conservative
bounds to the true values.

The variance, $\sigma^2$ is equal to $\sigma_{\rho}^2/\bar{\rho}^2$, where $\sigma_{\rho}^2$ is the 
variance in the mass density field, and $\bar{\rho}$ is its mean value.
A crude approach would be then to argue that 
\begin{equation}
\sigma = \frac{\sigma_{\rho}}{\bar{\rho}} \sim \frac {\sigma_{\rho}}{\rho} = \sigma_{\ln \rho},
\end{equation}
in which case $\sigma_{\log_{10} \rho}= \log_{10}(e) \sigma$. This would then mean that the
estimates for $\dot{\rho}_*$ in our lower (higher) redshift bins would have sampling errors
of $\pm0.52$dex ($\pm0.34$dex), respectively, so that the sampling variance in  $\dot{\rho}_*$
exceeds the uncertainty from the individual SFR estimates for both bins.

This method neglects the fact that, since $\sigma^2 \sim 1$, the probability distribution function
(PDF) for the cosmological density field will have been significantly skewed by gravitational
evolution. This skewness will be reflected in strongly asymmetric error bars on our $\dot{\rho}_*$ estimates
due to sampling variance: with such a skewed PDF, our randomly--selected survey volume (the selection of the HDF--S region was
based on the presence of a $z \sim 2$ quasar, which should have no bearing on the properties of galaxies
at $z \leq 1$) is much more likely to be underdense than overdense, and, hence, we are much more likely to have
measured a value of $\dot{\rho}_*$ that is lower than the global mean than one higher that it.
Estimating the level of such an effect
is very difficult. A full analysis would require a large numerical simulation of the cosmological density
field, coupled to a galaxy formation model capable of predicting accurately the sites of formation of
the class of galaxies detected in our {\em ISO\/} survey, which is far beyond the scope of this paper.
Simple analytic models prescriptions do exist for following the gravitational evolution of the PDF, but 
most are based on approximations ({\em e.g.\/} perturbation theory or the Zel'dovich approximation) that
break down by $\sigma \sim 1$. Bernardeau \& Kofman (1995) have, however, shown that a lognormal model
continues to give a good fit to the PDF derived from {\em N\/}--body simulations of a Cold Dark Matter universe
until at least $\sigma \sim 1.5$, so this is one analytic form that might be used. 

In Appendix B we show how this approximation may be used, within a Bayesian
framework, to compute the 68 per cent confidence intervals for $\bar{\dot{\rho}}_*$,
 the {\em global\/} star formation rate density
at the redshifts corresponding to the centres of our two bins, given our raw  $\dot{\rho}_*$ values and
our estimate that the matter variance in cells of size equal to our survey volumes is \mbox{$0.5 \leq \sigma^2 \leq
2$}. This analysis reveals that, for a best guess of $\sigma^2 =1$, 
\mbox{$0.62 \leq \bar{\dot{\rho}_*}/\hat{\dot{\rho}_*} \leq 3.2$}, while increasing or decreasing $\sigma^2$
by a factor of 2 barely changes the lower limit to $\bar{\dot{\rho}_*}/\hat{\dot{\rho}_*}$, while the upper
limit increases to 5.0 or decreases to 2.3, respectively. Thus we see that sampling variance introduces a
larger uncertainty into the estimation of  $\dot{\rho}_*$ than that caused by the uncertainties in the SFRs of
the individual galaxies in our survey. Assuming $\sigma^2 =1$, we estimate \mbox{ $0.02 \leq \bar{\dot{\rho}}_* \leq 0.10$} 
and \mbox{$0.04 \leq \bar{\dot{\rho}}_* \leq 0.23$} 
$M_{\odot} {\rm yr}^{-1} {\rm Mpc}^{-1}$ for the $z<0.43$ and $0.43 \leq z \leq 0.6$ bins, respectively, where we
define confidence intervals solely using our sampling variance analysis. 

We plot these confidence intervals on Fig.~\ref{fig:madau}.
These results are model--dependent, and neither the simple variance estimation nor the lognormal model of 
Appendix B is satisfactory, but both indicate that there is a large uncertainty associated with $\dot{\rho}_*$ 
estimates determined from volumes as small as our {\em ISO\/} survey of the HDF--S. Our results are consistent
with those of the {\em ISO\/}--based results of  Flores et al. (1999) (whose CFRS field is an order of magnitude larger, 
and so should yield $\dot{\rho}_*$ values with much lower sampling variances) and with the radio--based results of
Haarsma et al. (2000), but it is clear that, given the large sampling variance inevitable for such a small survey
volume, they cannot place tight constraints on the star formation history of the Universe by themselves.

\section{Discussion and Conclusions}\label{sect:discuss}

In this paper and its companion (Paper I) we have presented results from our {\em ISO\/} survey of
the Hubble Deep Field South. Here we sought optical identifications for the {\em ISO\/} sources
found in Paper I, obtaining reliable associations for 32 out of the 35 sources: these associations
should be more secure than those made by Mann et al. (1997) using our initial analysis of our
corresponding {\em ISO\/} survey of the northern Hubble Deep Field, thanks to the inclusion of
corrections for CAM image distortions, which were not well characterised in 1997. Of these 32
sources, a total
of twenty two were identified as spiral or starburst galaxies, eight of which have spectroscopic
redshifts (from the work of Rigpoulou et al. 2000 and Glazebrook et al. 2002) and the remaining
fourteen of which have had photometric redshifts estimated by ourselves and others. We estimate
that our photometric redshifts should be accurate to $\delta z =0.1$ or so, on the basis of the
set of eight objects for which spectroscopic redshifts are known, although we have noted
individual cases where the error is likely to be greater than that, for example when a redshift
has been estimated solely on the basis of JHK photometry. We found that the redshift distribution
of the galaxies associated with our {\em ISO\/} sources is consistent with that for similarly--bright
optical galaxies in the HDF--S region as a whole.

We reviewed a series of methods commonly used to determine star formation rates for galaxies
in  survey data, typically comprising a single broad--band flux for each object. We assessed
the ability of these to reproduce the SFRs of models of actively star--forming galaxies, finding,
as others (e.g. Granato et al. 2000) have, that those probing their far--infrared emission fared 
much better than those based on UV luminosity, indicating that massive star formation in these
galaxies takes place in dusty regions: the work of Rigopoulou et al. (2000) shows further that
this dust obscuration is not well corrected for using optically--determined extinction values. 
All these methods look for indications of the formation of massive stars, so a further complication
in the determination of the absolute rate at which stellar mass is being formed in a given galaxy
is uncertainty in the IMF, since the vast majority of the mass resides in stars not directly
probed by these methods, so SFR estimates differing by factors of a few can result if differing
IMFs are assumed. Caution must be exercised when applying these estimators in situations where
the SED type of the galaxies under study are constrained only by a small number of broad--band
fluxes, suggesting that constraints on the cosmic star formation history are best obtained
in well--studied fields, such as the Hubble Deep Fields, where rich, multiwavength datasets
are available. The small areas of the Hubble Deep Fields do, however, mean that significant
sampling variances exist for estimates of $\dot{\rho}_*$ at $z \leq 1$, and we showed, via two
simple methods for assessing their magnitude, that while our results are consistent with those
of previous authors, notably Flores et al. (1999) and Haarsma et al. (2000), they cannot by 
themselves yield tight constraints on the star formation history of the Universe, due to these
sampling effects.
Further details of this project can be found at {\tt astro.ic.ac.uk/hdfs}.

\section*{Acknowledgments}
This paper is based on observations with {\em ISO}, an ESA project, with
instruments funded by ESA Member States (especially the PI countries:
France, Germany, the Netherlands and the United Kingdom) and with
participation of ISAS and NASA.
This work was in part supported by PPARC grant no.  GR/K98728
and  EC Network is FMRX-CT96-0068. We thank the ATNF HDF--S team, particularly Andrew Hopkins,
for providing us with radio data for our sources in advance of publication, and an anonymous
referee for comments.

\appendix

\section{Conversion of SFR estimators to a canonical IMF}

As mentioned in Section 3.2, there are three types of physical model underlying the SFR estimators of
Eqns. (2) - (8). It follows, by the definition of $\dot{M}$, the rate of formation of stellar mass
 (in $M_{\odot} {\rm yr}^{-1}$), that
\begin{equation}
\dot{M} = \int^{M_U}_{M_L} M \psi(M) {\rm d}M,
\label{eq:mdot}
\end{equation}
where $\psi(M)$ is the IMF , and stars are being formed over the mass range $[M_L,M_U]$.

The first class are those (from Eqns. 5, 6 \& 7) based on the assumption (Thronson \& Telesco 1986,
Rowan--Robinson et al. 1997) that some large ($\epsilon\simeq 1$) fraction of the bolometric luminosity, 
$L_{\rm bol}$, generated in any burst of star formation emerges in the far infrared (FIR), due to the reprocessing 
by dust of the light from young stars: {\em i.e.\/}
\begin{equation}
L_{\rm FIR} = \epsilon L_{\rm bol} = \epsilon \int^{M_U}_{M_L} L(M) t_{\rm FIR}(M) \psi(M) {\rm d}M,
\end{equation} 
where $L(M)$ is the luminosity of a newly--formed star of mass $M$ (and it is assumed that only main sequence
stars contribute) and that such stars contribute to $L_{\rm FIR}$ for time $t_{\rm FIR}(M)$. 
The crucial timescale here is the time taken for the dust cloud around the young star to be destroyed and that
is assumed (Thronson \& Telesco 1986) to be independent of mass, and to be $\tau_{\rm FIR}\sim O(10^6$ yr),
in which case
\begin{equation}
L_{\rm FIR} =  \epsilon \tau_{\rm FIR} \int^{M_U}_{M_L} L(M) \psi(M) {\rm d}M,
\label{eq:lfir}
\end{equation}
so that combining Eqns.~\ref{eq:mdot} and~\ref{eq:lfir} 
\begin{eqnarray}
\dot{M} & \propto & L_{\rm FIR} \cdot \frac {\int^{M_U}_{M_L} M \psi(M) {\rm d}M}{\int^{M_U}_{M_L} L(M) \psi(M) {\rm d}M} \nonumber \\
& \propto & L_{\rm FIR} \frac{\bar{M}}{\bar{L}_{\rm bol}},
\end{eqnarray}
where $\bar{M}$ and $\bar{L}_{\rm bol}$ are, respectively, the average 
mass and bolometric luminosity of stars formed according to the particular IMF: {\em i.e.\/}
\begin{equation}
\bar{M} = \frac {\int^{M_U}_{M_L} M \psi(M) {\rm d}M}{\int^{M_U}_{M_L}  \psi(M) {\rm d}M}
\end{equation}
 and
\begin{equation}
\bar{L} = \frac {\int^{M_U}_{M_L} L(M) \psi(M) {\rm d}M}{\int^{M_U}_{M_L}  \psi(M) {\rm d}M}.
\end{equation}
So, we have shown, following Thronson \& Telesco (1986) that the SFR (in $M_{\odot} {\rm yr}^{-1}$) per unit of far 
infrared luminosity (say, at 60$\mu$m, or integrated over 3--1000$\mu$m) is proportional to $\bar{M}/\bar{L}_{\rm 
bol}$. 

Several choices exist for the form of $L(M)$ to use. Telesco \& Gatley (1984) assume a double power law
form, namely $L(M)/L_{\odot}= A \cdot (M/M_{\odot})^\alpha$, where
\begin{equation}
(A,\alpha) = \left\{   \begin{array}{ll} (1.3, 3.6)  \; 0.1 \leq M/M_{\odot} \leq 10 \\
 					 (8.1, 2.8)  \;\;  10 \leq M/M_{\odot} \leq 60,
		       \end{array} \right.
\label{eq:lm_tel-gat}
\end{equation}	
using which yields the following set of $\bar{M}/\bar{L}_{\rm bol}$ values for the $\psi(M) \propto M^{-x}$ models 
of interest here:
\begin{eqnarray}
&  &\frac{ \bar{M}/\bar{L}_{\rm bol} }{ 10^{-3}\bar{M}_{\odot}/\bar{L}_{\odot} }  = \nonumber \\
&  & \left\{   \begin{array}{lll} 
\!\!\!\! 1.33:    \; (M_L, M_U)=(0.1,100) M_{\odot}, x=2.35 \\
\!\!\!\!\!\! 1.02:    \; (M_L, M_U)=(0.1,120) M_{\odot}, x=2.35 \\
\!\!\!\! 2.52:    \; (M_L, M_U)=(0.1,100) M_{\odot}, x=2.50, 

		       \end{array} \right.
\label{eq:m-over-l}
\end{eqnarray}
reproducing the results in Table 3 of Thronson \& Telesco (1986), from which Rowan--Robinson et al. (1997) 
derived the value that $\phi$ should take in eqn (5) for different choices of IMF. Following this method, we 
find that, to match our reference IMF, requires having $\phi=1$ in equation (5), and multiplying the SFR deduced
from eqn. (7) by a factor of $(1.33/1.02)^{1.05}=1.32$ to account for the fact that Devriendt et al. (1999)
use a Salpeter IMF with an upper mass limit of 120$M_{\odot}$.

To calculate the scaling appropriate for the 60$\mu$m estimator of Cram et al. (1998), we must not only account 
for the different $\bar{M}/\bar{L}_{\rm bol}$ values for their IMF and our reference Salpeter (1955) law (given 
above), but also the fact that they quote an SFR which refers to stars of mass $M \geq 5 M_{\odot}$, which is done 
as follows: only one ninth of the
stellar mass formed with their IMF is in stars of $M \geq 5 M_{\odot}$, with the result that the SFR
value from equation (6) must be multiplied by a factor \mbox{$9/(2.52/1.33)=4.8$} to make it appropriate for the
formation of stars according to our reference IMF. 

The second class of methods comprises the three SFR estimators based on monochromatic UV luminosities. In
this case, the factors of  $\bar{M}/\bar{L}_{\rm bol}$ in the scaling are replaced by 
$\bar{M}/\bar{L}_\lambda$, where $\bar{L}_\lambda$ is the mean value of the particular monochromatic
luminosity under consideration, evaluated over the stars formed according to the given IMF. Values for
$\bar{L}_\lambda$ can be readily computed under the assumption that main sequence stars emit as black
bodies in the ultraviolet. This we implement, making double and triple power law fits, respectively, to
the main sequence radius--mass and effective temperature--mass relation data tabulated by Binney \& Merrifield
(1998), {\em i.e.\/}:
\begin{eqnarray}
& & \log_{10}\left( \frac {R}{R_{\odot}} \right) = \nonumber \\
& & \left\{   \begin{array}{ll} 
 					0.02+0.72 \log_{10} (M/M_{\odot}):  \; M/M_{\odot} \leq 10 \\
					0.30+0.44 \log_{10} (M/M_{\odot}):  \; M/M_{\odot} \geq 10
\end{array} \right.
\label{eq:bm_r-m_fit}
\end{eqnarray}	
and
\begin{eqnarray}
\label{eq:bm_t-m_fit}
& & \log_{10}\left( \frac {T_{\rm eff}}{10^3 \; {\rm K}} \right) =   \\
& & \left\{   \begin{array}{lll} 
\!\!\!0.63+0.17 \log_{10} (M/M_{\odot}):  \; M/M_{\odot} \leq 0.5 \\
\!\!\!0.76+0.59 \log_{10} (M/M_{\odot}):  \; 0.5 \leq M/M_{\odot} \leq 1.6\\
\!\!\!1.11+0.30 \log_{10} (M/M_{\odot}):  \; M/M_{\odot} \geq 1.6.
 							       \end{array} \right. \nonumber
\end{eqnarray}	
(Note that, if we use these scalings and the relation $L_{\rm bol}/L_{\odot}=\left(R/R_{\odot}\right)^2 
\left(T_{\rm eff}/5770{\rm K}\right)^4$, instead of the $L(M)$ relation of Eqn.~\ref{eq:lm_tel-gat}, we
obtain ratios of $\bar{M}/\bar{L}_{\rm bol}$ for pairs of IMF models which differ by $\sim$20\%, 
typically, from those of Eqn.~\ref{eq:m-over-l}.) $\bar{L}_{\lambda}$ values may then be computed, using the
fact that $L_{\lambda}=\int l_{\lambda}(M) \psi(M) {\rm d}M$, where $l_{\lambda}(M)$, the monochromatic 
luminosity at wavelength $\lambda$ due to stars of mass $M$ is given by $l_{\lambda}=4 \pi R(M)^2 B_{\lambda}
[T_{\rm eff}(M)]$, where $B_{\lambda}$ is the Planck function, and $R$ and $T_{\rm eff}$ come from Eqns.
~\ref{eq:bm_r-m_fit} and ~\ref{eq:bm_t-m_fit}.

For the two UV--based estimators of Devriendt et al. (1999) we must make conversions at $\lambda$ =
160 and 280nm, from a Salpeter (1955) IMF with mass range $[0.1,120]M_{\odot}$ to our canonical 
range of $[0.1,100]M_{\odot}$. The ratio of the $\bar{M}$ values for the two IMFs is unity to better
than 1\% precision, while the  $\bar{L}_{\lambda}$ values differ by $\sim10$\%: we obtain scalings of
 $1.1^{1.62}$ and $1.1^{1.72}$, respectively, for the  280nm and 160nm estimators, accounting for the
non--linear relationship between $\bar{L}_{\lambda}$ and the SFR from Devriendt et al. (1999). As
before, we must account for the fact that the Cram et al. (1998) U band estimator refers to the
formation of stars of mass greater than $5 M_{\odot}$ only, when we derive its correction factor,
as well as considering the difference between the $\bar{L}_{\lambda}$ value at 250nm between the
Cram et al. (1998) IMF and our canonical model. $\bar{M}/\bar{L}_\lambda$ for the Cram et al. (1998)
IMF is only 57 per cent of that for the canonical IMF, reflecting their difference in slope, so, to
correct the SFR estimator of eqn (2) we must multiply it by \mbox{$0.57 \times 9=5.1$}.

The 1.4 GHz estimator of Cram et al. (1998) constitutes the third class, and it is different in that
it relates the SFR to the {\em number\/} of stars formed, rather than to their luminosity. As 
presented by Condon (1992), the argument for using decimetric radio luminosity as a measure of star
formation rate hinges on the assumption that the radio emission from active star--forming galaxies
is dominated by non--thermal emission from the remnants of Type Ib and Type II supernovae. In this
picture, the radio luminosity is, thus, proportional to the (Type Ib and Type II) supernova rate,
$\nu_{\rm SN}$, which is equated to the rate of formation of progenitors of sufficient mass ($M \geq M_0$):
\begin{equation}
\nu_{\rm SN} = f_{\rm N}(M \geq M_0) \times (\dot{M}/\bar{M}),
\end{equation}
where $f_{\rm N}(M \geq M_0)$ is the fraction of stars formed which have mass $M \geq M_0$, and
$\dot{M}/\bar{M}$ equals the number of stars formed per unit time. Since, the decimetric radio luminosity, 
$L_{\rm cm}$, is proportional to $\nu_{\rm SN}$, it follows that 
\begin{equation}
\dot{M} \propto \left( \frac {\bar{M}}{f_{\rm N}(M\geq M_0)} \times L_{\rm cm} \right),
\end{equation}
which gives the correct scaling of $\dot{M}$ per unit radio luminosity between IMFs. If we follow
Condon (1992) and take $M_0=8 M_{\odot}$, we find that
the $\bar{M}/f_{\rm N}(M\geq M_0)$ value for our canonical IMF is 65 per cent that for the IMF
assumed by Cram et al. (1998), reflecting smaller fraction of high--mass stars in the $M^{-2.5}$
IMF, and from this we find that the factor by which the SFR given in eqn (8) has to be multiplied 
is $0.65 \times 9=5.8$, where, as before, the factor of nine corrects for the fact that eqn (8) 
only gives the rate of formation of stellar mass in stars of mass greater than 5 $M_{\odot}$. 

Similar procedures must be applied to convert the SFR estimates presented (for a Salpeter IMF
over the mass range [1,100]$M_{\odot}$) by Rigopoulou et al. (2000) to our canonical IMF: for their
far--infrared estimator, this is an $\bar{M}/\bar{L}_{\rm bol}$ scaling, while for their H$\alpha$
estimator, we must scale by $\bar{M}/\bar{L}_{\lambda}$, for $\lambda=6560 \AA$. Coincidentally, we
find that, in both cases, this yields a factor of 2.5 by which their SFR values must be multiplied
for comparison with others computed for our canonical IMF. A further correction, however, must be applied,
since Rigopoulou et al. (2000) assumed a cosmology with $\Omega_{\rm M}=0.3$ and $\Omega_{\Lambda}=0.7$,
which yields a larger luminosity distance out to a given redshift than does our canonical Einstein -- de Sitter 
model. At the typical redshift ($z \sim 0.6$), this reduces the SFR estimates by $\sim 30$ per cent, so that
the Rigopoulou et al. (2000) SFRs are to be multiplied by typical values of 1.25 before comparison with our own.

\section{A lognormal model for the sampling bias and variance in $\dot{\rho}_*$}

We want to compute confidence ranges for $\bar{\dot{\rho}_*}$, the true, global mean SFR density at $z\sim 0.2$ and
$z \sim 0.5$ on the basis of our raw $\dot{\rho}_*$ values, given in Section~\ref{sect:raw_madau}. 
An accurate determination of these confidence ranges
requires a model for the galaxy population at $0 \leq z \leq 1$ of a sophistication
that is beyond the scope of this paper. However, we may obtain an estimate for them 
as follows.

If we denote the
SFR density by $X$, so that the desired global mean is $\bar{X}$ and an estimate is 
denoted by $\hat{X}$, then we wish to determine the probability distribution $p(\bar{X}|\hat{X})$.
We may adopt a Bayesian approach, writing
\begin{equation}
p(\bar{X}|\hat{X}) = \frac{p(\bar{X},\hat{X})}{p(\hat{X})} = \frac{p(\hat{X}|\bar{X}) \cdot p(\bar{X})}
{p(\hat{X})}.
\end{equation}
Now, if we assume a uniform prior for $p(\bar{X})$ we find that the relative probability of 
$\bar{X}$ given $\hat{X}$ is
\begin{equation}
p(\bar{X}|\hat{X}) \propto p(\hat{X}|\bar{X}) = p(1+\delta),
\end{equation}
where $\delta$ is the fractional density fluctuation.  
Since our estimates from Section~\ref{sect:sampling} for the variances in our two sampling volumes are
of order unity, we cannot assume that the distribution of 
$\delta$ is Gaussian. Indeed, gravity will have skewed the distribution so that it is 
significantly more likely that a randomly--located survey volume will be underdense than overdense,
so we are more likely to have under--estimated $\bar{\dot{\rho}_*}$ than to have over--estimated it.
For present purposes, we assume
a lognormal model for the density field, because Bernardeau \& Kofman (1995) argue 
that it gives a good fit to the probability density function of the density fluctuations in 
{\em N\/}--body simulations of a Cold Dark Matter universe well into the non--linear
regime (i.e. at least until the variance in fractional overdensity reaches $\sim 1.5^2$).

\begin{figure}
\hspace{-0.7cm}
\epsfig{file=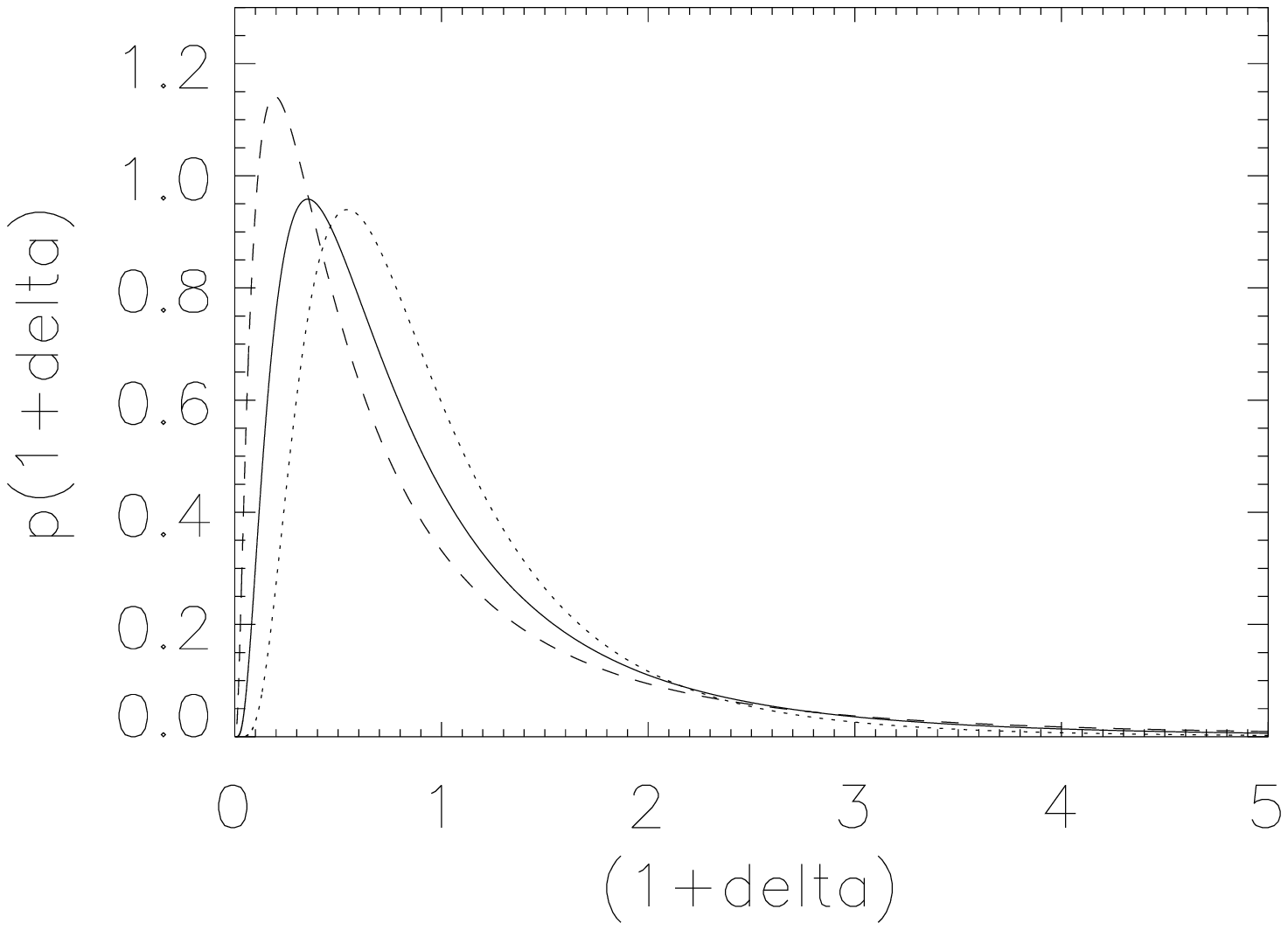,width=8cm,angle=0}
\caption{The probability density functions $p(1+\delta)$ as a function of $1+\delta \equiv \hat{X}/\bar{X}$
in the lognormal model for the three cases of $\sigma^2=$0.5, 1.0 \& 2.0, plotted with dotted, solid and dashed lines
respectively.}
\label{fig:evolved_pdf}
\end{figure}

In the lognormal model, we have
\begin{eqnarray}
& & p(1+\delta) {\rm d}(1+\delta)  =   \frac{1}{\sqrt{(2 \pi \sigma_0^2)}}  \nonumber \\ 
& & \times  \exp \left[ - 
\frac{(\ln (1+\delta) + \sigma_0^2/2)^2} {2 \sigma_0^2} \right] \frac { {\rm d}(1+\delta)}{(1+\delta)},
\label{eqn:lognormal}
\end{eqnarray}
where $\sigma_0^2=\ln(1+\sigma^2)$ is the variance of the fictitious Gaussian field from which the
true density field, with variance $\sigma^2$, is supposed to be derived, by taking the natural logarithm. It can
readily be shown that \mbox{$p(1+\delta)$} takes its maximum value at $1+\delta= \exp (-3 \sigma_0^2 /2)$,
where $p(1+\delta) = \exp (\sigma_0^2) /\sqrt{(2 \pi \sigma_0^2)}$. 

In Fig.~\ref{fig:evolved_pdf} we plot $p(1+\delta)$ against  $1+\delta \equiv \hat{X}/\bar{X}$
for the three cases of $\sigma^2=0.5$ (dotted line), 1.0 (solid line)  and 2.0 (dashed line), assuming the lognormal
model of eqn.~\ref{eqn:lognormal}. The relatively small fraction of area under the curves at $1+ \delta > 1$,
indicates the low probability that the true global mean SFR density will be lower than that
estimated, reflecting the fact that for $\sigma^2 \geq 0.5 $ much more of the universe is underdense than overdense.
However, the mean value of $1+ \delta$ must always equal unity, by definition, so our estimated SFR density is
not a biased estimate, but the evolution of the cosmological density field through gravitational instability does
act to increase the width of the confidence range for  $\bar{\rho}_*$ at a given percentile. 

For the lognormal model, this is simple is to compute. To obtain the width of, say, the 68 per cent confidence region
for $1+\delta$, we want to take the difference $z_2 - z_1$ where $z \equiv 1 + \delta=z_1$ and $z=z_2$ are the solutions 
to the equation 
\begin{equation}
\int_0^{z} p(1+\delta) {\rm d}(1+\delta)  = \tilde{p}
\end{equation}
for $\tilde{p}$=0.16 and 0.84 respectively. It is easy to show that 
\begin{equation}
\int_0^{z} p(1+\delta) {\rm d}(1+\delta)  = \frac{1}{2} \left[ 1 + {\rm erf} \left( \frac { \ln(z) + \sigma_0^2 /2}
{\sqrt{2 \pi \sigma_0^2}} \right) \right], 
\end{equation}
where ${\rm erf}(x)$ is the error function of $x$, and, hence, to obtain 68 per cent confidence ranges of
\mbox{$0.65 \leq \bar{X}/\hat{X} \leq 2.3$}, \mbox{$0.62 \leq \bar{X}/\hat{X} \leq 3.2$}, and 
\mbox{$0.61 \leq \bar{X}/\hat{X} \leq 5.0$}, for $\sigma^2$=0.5, 1.0 and 2.0, respectively. It is interesting to
note that, in this lognormal model, at least, the effect of continued evolution of the density field
via gravitational instability is really only seen in the shift of the upper limit to the confidence interval for
$\bar{X}/\hat{X}$, and that the lower limit barely changes: so, as time passes and $\sigma^2$ increases, the amount
by which one's survey may have under--estimated the true, global mean SFR density increases markedly, but the 
amount by which it may have over--estimated it stays pretty much the same. This is equivalent to the noting that the difference
between the three curves in Fig.~\ref{fig:evolved_pdf} is much more pronounced at low values of $1+\delta$, while,
at high values, they are very close.

\label{lastpage}

\end{document}